\documentclass[12pt]{article}

\RequirePackage{amsthm,amsmath}
\RequirePackage[round]{natbib}
\RequirePackage{amssymb}
\usepackage{mathrsfs}
\usepackage{mathtools}
\usepackage{bm}
\usepackage{amsfonts}
\usepackage{bbm}
\usepackage{mathrsfs} 
\usepackage{booktabs}
\usepackage{graphicx}
\usepackage{chngpage}
\usepackage{multirow}
\usepackage{nicefrac}
\usepackage[section]{placeins}
\usepackage{soul}
\usepackage[export]{adjustbox}
\usepackage{enumitem}
\usepackage[T1]{fontenc}
\usepackage[utf8]{inputenc}
\usepackage{authblk}
\usepackage{ulem}

\usepackage{wrapfig}
\usepackage{lscape}
\usepackage{rotating}
\usepackage{comment}
\usepackage{subfig}

\usepackage{tikz}
\usepackage{color}
\usetikzlibrary{positioning}
\usetikzlibrary{arrows}

\usepackage{mathtools}

\usepackage{xr-hyper}
\usepackage{pdfpages} 
\usepackage[colorlinks=true, linkcolor=red, urlcolor=blue, citecolor=blue]{hyperref}

\makeatletter
\newcommand*{\addFileDependency}[1]{
  \typeout{(#1)}
  \@addtofilelist{#1}
  \IfFileExists{#1}{}{\typeout{No file #1.}}
}
\makeatother
 
\newcommand*{\myexternaldocument}[1]{%
    \externaldocument{#1}%
    \addFileDependency{#1.tex}%
    \addFileDependency{#1.aux}%
}

\myexternaldocument{rgmwm_app}

\usepackage{setspace}
\allowdisplaybreaks
\definecolor{BRC}{rgb}{0.0, 0.26, 0.15}
\definecolor{carib}{rgb}{0.0, 0.8, 0.6}
\definecolor{darkpastelgreen}{rgb}{0.01, 0.75, 0.24}
\definecolor{burgundy}{rgb}{0.5, 0.0, 0.13}

\def\boxit#1{\vbox{\hrule\hbox{\vrule\kern3pt
          \vbox{\kern3pt#1\kern3pt}\kern3pt\vrule}\hrule}}

\DeclareMathOperator*{\argzero}{argzero}

\DeclareMathOperator*{\med}{med}
\DeclareMathOperator*{\mad}{mad}
\let\iid\undefined
\DeclareMathOperator{\iid}{\textup{iid}}

\DeclareSymbolFont{lettersA}{U}{txmia}{m}{it}
\DeclareMathSymbol{\real}{\mathord}{lettersA}{"92}
 
\numberwithin{equation}{section}
\theoremstyle{plain}
\newtheorem{thm}{Theorem}[section]
\newtheorem{lemma}{Lemma}[section]

\newtheorem{remark}{Remark}[section]

\newtheorem{prop}{Proposition}[section]

\newcommand{\blind}{0}

\begin{document}

\def\spacingset#1{\renewcommand{\baselinestretch}%
{#1}\small\normalsize} \spacingset{1}


\if0\blind
{
  \title{\bf Robust Two-Step Wavelet-Based Inference \\for Time Series Models}
  \author{
  St\'ephane Guerrier\\
  \vspace{-0.4cm}
  GSEM, University of Geneva, Switzerland\\
  \vspace{0.1cm}
  Roberto Molinari\\
  Pennsylvania State University, USA\\
  \vspace{0.1cm}
  Maria-Pia Victoria-Feser\\
  GSEM, University of Geneva, Switzerland\\
  and\\
  \vspace{0.1cm}
  Haotian Xu\\
  GSEM, University of Geneva, Switzerland}
  \date{}
  \maketitle
} \fi

\if1\blind
{
  \bigskip
  \bigskip
  \bigskip
  \begin{center}
    {\LARGE\bf Robust Two-Step Wavelet-Based Inference \\for Time Series Models}
\end{center}
 \medskip
} \fi

\bigskip

\begin{abstract} 


Complex time series models such as (the sum of) ARMA$(p,q)$ models with additional noise, random walks, rounding errors and/or drifts are increasingly used for data analysis in fields such as biology, ecology, engineering and economics where the length of the observed signals can be extremely large. Performing inference on and/or prediction from these models can be highly challenging for several reasons: (i) the data may contain outliers that can adversely affect the estimation procedure; (ii) the computational complexity can become prohibitive when models include more than just a few parameters and/or the time series are large; (iii) model building and/or selection adds another layer of (computational) complexity to the previous task; and (iv) solutions that address (i), (ii) and (iii) simultaneously do not exist in practice. For this reason, this paper aims at jointly addressing these challenges by proposing a general framework for robust two-step estimation based on a bounded influence M-estimator of the wavelet variance. In this perspective, we first develop the conditions for the joint asymptotic normality of the latter estimator thereby providing the necessary tools to perform (direct) inference for scale-based analysis of signals. Taking advantage of the model-independent weights of this first-step estimator that are computed only once, we then develop the asymptotic properties of two-step robust estimators using the framework of the Generalized Method of Wavelet Moments (GMWM), hence defining the Robust GMWM (RGMWM) that we then use for robust model estimation and inference in a computationally efficient manner even for large time series. Simulation studies illustrate the good finite sample performance of the RGMWM estimator and applied examples highlight the practical relevance of the proposed approach.
\end{abstract}

\noindent%
\vfill
 
\newpage
\spacingset{1.5} 
\section{Introduction}
\label{sec.intro}

As for many other fields of statistical research, time series analysis is also facing different challenges due to the increasing amounts of data being recorded over time within a wide variety of contexts going from biology and ecology to finance and engineering. Among others, these challenges include (i) the need for more complex (parametric) models (e.g. to deal with possible long-term non-stationary features), (ii) the need for computationally efficient (or simply feasible) methods to analyze data and estimate such models as well as (iii) the need to deal with the increased probability of observing measurement errors in the form of contamination (outliers), model deviations, etc. In this context, there is currently a large variety of models and statistical methods available to analyze and draw conclusions from time series \citep[see e.g.][for an overview]{percivalbook,durbin2012time, shumway2013time}. However, many of the existing methods may become unfeasible, for example, when estimating even slightly complex models on large time series data. In addition, the use of robust methods to perform inference when the data suffers from contamination is often a daunting task even for relatively simple (parametric) settings and moderate sample sizes, without considering more complex models in larger data settings \citep[for motivating examples and robust inferential approaches, see e.g.][]{MaMaYo:06,maronna2019robust}.

In response to the above challenges and limitations, this paper proposes an alternative and general robust inference framework for (latent) parametric time series models that include (the sum of) ARMA and rounding error models as well as non-stationary models such as drifts and random walks. This framework is based on a two-step approach where the first (and most important) step consists in the proposal of a robust estimator of the Wavelet Variance (WV) \citep[see e.g.][]{percival1995estimation} with adequate asymptotic properties based on minimal conditions, while the second step integrates these results within the Generalized Method of Wavelet Moments (GMWM) framework \citep[see][]{gmwm} for the purpose of inference on time series models\footnote{The method (with relative graphical tools) is implemented partly in the \texttt{simts} R package on CRAN and in a more complete manner in an open source R package, available at \url{github.com/SMAC-Group/gmwm}; see also \citet{GMWM-online-2018,BaRaNaGuZhMo:18,RaBaGuESSeMo:19}}. With regards to the first step, the WV has been widely used within the natural and physical sciences for analysis of variance, model building and prediction \citep[see e.g.][for an overview]{percivalbook}, and more recently within other fields of research \citep[see e.g.][to mention a few]{gallegati2012wavelet, xie2013wavelet,foufoula2014wavelets, jia2015correlations, ziaja2016fault, abry2018wavelet}. In this context, the properties of the standard estimators of WV have been developed \citep[see][]{percival1995estimation, serroukh2000statistical} while their limitations in presence of contamination or outliers have been underlined in \citet{mondal2012mest} where a robust estimator was also put forward for this purpose. However, as explained further on, the use of an alternative M-estimator is preferable in different settings, including the two-step framework considered in the present paper.

With respect to parametric inference for time series, robust methods (including two-step approaches) have been proposed in abundance over the past decades and a detailed overview of these can be found, for example, in \citet{MaMaYo:06, maronna2019robust} (Chapter 8) while Appendix~\ref{LitRev} also provides a short literature review. More specifically, a general approach in this context is to maximize \textit{asymptotic} efficiency of the resulting estimators with respect to a specific model (or class of models). For this purpose, a traditional approach consists in placing estimation (and inference) in the framework of bounded-influence estimators that are obtained by bounding the corresponding estimating equations \citep{HaRoRoSt:86}, such as the Maximum Likelihood Estimator (MLE) equations or indeed those of other efficient (non robust) estimators. In this setting, a correction factor is often required to ensure consistency of the resulting robust estimator and needs to be defined for each model considered for a given time series. The computation of this model-dependent correction factor generally requires numerical approximations of possibly high-dimensional integrals or multiple one-dimensional integrals (for conditionally unbiased estimators, see \citealp{Kuns:84}), whose dimensions or number  grow proportionally to the sample size. These consistency corrections can therefore not only be too burdensome to compute,  relying on more computationally intensive simulation-based procedures to approximate these quantities, but also are rarely accounted for in the derivation of the statistical properties of the resulting estimator.
On the other hand, robust two-step estimators (based for example on robust filtering methods or on robust autocovariance/autocorrelation estimators), while possibly paying a price in terms of asymptotic efficiency, deliver not only computational advantages but also allow to compare candidate models under a different perspective.  Indeed, the first-step estimators provide a set of estimates that remain common to all candidate models and have a bounded  Influence Function (IF) \citep{Hamp:74} through weights (for weighted estimators) that are computed \textit{only once} independently from the candidate models. Using these common estimates, aside from guaranteeing that the second step estimators inherit their bounded IF property (hence robustness), the consequence is that the model comparison in terms of fit (or prediction accuracy) is made solely on the structural features of the candidate models. This provides an alternative approach to model comparison based on robust estimators since, in the weighted estimating equation setting, the estimation weights are relative to each model under consideration that is, by assumption, the correct one. On the other hand, a two-step approach alleviates the latter hypothesis in the computation of the robust weights and hence provides an alternative route to robust model comparison. Moreover, this allows to develop graphical robust model selection tools which, in the context of this paper, rely only on a few quantities (even for large samples as seen further on and in Section \ref{sec:applications}) as well as to build robust model selection criteria in a relatively straightforward manner.

However, while robust two-step approaches have noticeable computational and model comparison advantages, these can nevertheless suffer from some theoretical and practical drawbacks. Indeed, when they are based on robust filtering, they may lead to biased estimators and lack asymptotic theory in order to perform adequate inference \citep[see e.g.][]{MuPeYo:09}. Alternatively, in the case where two-step estimators are based on robust moment estimators (e.g. autocovariance or autocorrelation), these are often computed in the framework of indirect inference \citep{smith1993estimating,gourieroux1993indirect} or, similarly, in that of the (simulated) method of moments \citep{mcfadden1989method, gallant1996moments, duffie1993simulated}. In these settings, two-step estimators can become computationally intensive in large samples since the number of auxiliary moments that are available in practice increase linearly with the sample size, thereby requiring a moment selection procedure which delivers additional uncertainty in the successive inference phase \citep[see e.g.][]{andrews1999consistent}. However, the GMWM framework considerably reduces the latter drawback since its first-step moment is the WV which adequately summarizes the information in the spectral density (or autocovariance) function into $J < \log_2(T)$ moments (with $T$ representing the sample size) thereby greatly reducing the number of moments even for large sample sizes while also allowing for inference on non-stationary models. Being a natural extension of the GMWM framework, the Robust GMWM (RGMWM) relies on the proposed robust estimator of WV and makes use of its properties in order to derive the inferential properties of the RGMWM so as to perform adequate estimation and general inference procedures for a wide range of time series models under slight model deviations (such as the presence of outliers) in a computationally efficient manner. 

Considering the above goal, the paper is organized as follows. Section~\ref{sec:wv} proposes the robust WV estimator that constitutes the first step of the RGMWM that is presented in Section~\ref{sec:rgmwm}. In both sections, the necessary asymptotic results are developed in order to perform robust estimation and inference for (intrinsically) stationary time series. Section~\ref{sec:simulations} presents a simulation study where the robustness properties of the RGMWM estimator are compared with other estimators and Section~\ref{sec:applications} demonstrates the practical usefulness of the proposed method in an economic setting while another applied setting is presented in Appendix~\ref{iner.sens}.

\section{Robust Wavelet Variance}
\label{sec:wv}


To introduce the WV, let us denote $(X_t)$, $t=1,\hdots,T$, as an \textit{intrinsically} stationary time series that is either stationary or becomes so when a d$^{th}$-order backward difference is applied to it. Moreover, let 
\begin{equation*}
	W_{j,t} := \sum_{l = 0}^{L_j - 1} h_{j,l} X_{t - l} \, ,
\end{equation*}
denote the wavelet coefficients that result from a wavelet decomposition of the time series, where $(h_{j,t})$ is a known wavelet filter of length $L_j$ at (dyadic) scale $\tau_j$, for $j = 1, \hdots, J$ and $J < \log_2(T)$. Since the filter sequence $(h_{j,t})$ has specific properties, the wavelet coefficients can be seen as the result of a particular form of weighted moving-average taken over a number of observations proportional to $L_j$ \citep[which increases as $j$ increases, see e.g.][for a general overview]{percivalbook}. Moreover, if $L_1 \geq \text{d}$ and $(X_t)$ is a d$^{th}$-order stationary time series, the resulting wavelet coefficients will be stationary. Based on the (stationary) wavelet coefficients, the WV is defined as
\begin{equation*}
\nu_{j}^2 := \text{Var}(W_{j,t}) ,
\end{equation*}
i.e. the variance of the wavelet coefficients, which can be expressed in vector form as $\bm{\nu} := [\nu_j^2]_{j = 1, \hdots, J}$. While this definition holds independently from a possible parametric model underlying the time series $(X_t)$, for the purpose of this work let us assume that the latter is generated from the parametric family of models $F_{\bm{\theta}_0}$, with unknown parameter vector denoted by $\bm{\theta}_0 \in \bm{\Theta} \subset \mathbb{R}^p$. In order to make the link between the WV and an assumed \textit{stationary} parametric model $F_{\bm{\theta}}$ explicit, an alternative representation of this quantity is based on the Spectral Density Function (SDF) and is as follows:
\begin{equation*}
	\nu_{j}^2 (\bm{\theta}) = \int_{-1/2}^{1/2} |H_j(f)|^2 S_{F_{\bm{\theta}}}(f) df \, ,
\end{equation*}
with $S_{F_{\bm{\theta}}}(f)$ denoting the theoretical SDF and $H_j(f)$ being the Fourier transform of the wavelet filters $(h_{j,t})$. The above equality is valid also when considering the theoretical spectral density of the wavelet coefficients themselves which, depending on the length of the wavelet filter, is defined even if the original time series $(X_t)$ is not stationary \citep[see e.g.][]{percivalbook}. Although the results of the present section hold without any parametric assumption for $(X_t)$, in this paper we will generally assume that the true WV will depend on a parametric family of models $F_{\bm{\theta}}$ and use the notation $\bm{\nu}$ for this purpose. We will make use of the notation $\bm{\nu}(\bm{\theta})$ whenever it is appropriate to explicitly highlight the link with the underlying parametric model.

Based on the above definitions, each parametric time series model has a corresponding theoretical WV vector which summarizes the information contained in the SDF or AutoCovariance Function (ACF) into few quantities ($J < \log_2(T)$) for a wide range of intrinsically stationary time series models (e.g. SARIMA and many state-space models). Considering this, when compared to the SDF or ACF, the information loss of the WV can be offset by its ability to adequately represent a considerably wide range of stationary and non-stationary time series. In this sense, aside from being discussed in \citet{greenhall1998spectral} and \citet{gmwm}, Appendix~\ref{app_ident} includes additional results proving that the WV is capable of adequately recovering information on time series models thereby supporting the usefulness of this quantity.

For the rest of this section however, let us disregard any parametric assumption for $(X_t)$ and simply assume that its corresponding wavelet coefficients are stationary. With this in mind, several estimators have been proposed for the WV, the main one being the standard unbiased estimator of the WV proposed by \citet{percival1995estimation} and defined as
\begin{equation}
    \tilde{\nu}_{j}^2 := \frac{1}{M_{j}} \sum_{t=1}^{M_j} W_{j, t}^2 \,\,,
    \label{eq:standard_wvest}
\end{equation}
where $M_j$ is the length of the wavelet coefficient process $(W_{j, t})$ at scale $\tau_j$. The theoretical properties of this estimator were further studied in \citet{serroukh2000statistical} in which the conditions for its asymptotic properties are given. However, as highlighted by \citet{mondal2012mest}, the estimator of WV in (\ref{eq:standard_wvest}) is not robust and can therefore become considerably biased in the presence of outliers or different forms of data contamination. For this reason they put forward a robust estimator of the WV (developed for Gaussian time series specifically affected by scale-based contamination) by making use of a log-transformation of the (squared) wavelet coefficients to apply standard M-estimation theory for location parameters. However, due to these transformations and the approximate corrections to reverse them, the asymptotic properties of the final estimator are not straightforward to define. Therefore, considering the robustness issues of the standard estimator $\tilde{\nu}_{j}^2$ and the ``limitations'' of the latter robust estimator, the following paragraphs put forward a new M-estimator of WV that overcomes these issues and, as shown in Appendix~\ref{sim_wv}, performs generally better in finite sample settings.

\subsection{M-Estimation of Wavelet Variance}
\label{sec:wv_mest}

Following the above discussion, this section generalizes the standard estimator of WV proposed in \citet{percival1995estimation} to an M-estimator \citep{Hube:64} which can also be made robust by choosing a bounded score function, thereby delivering an appropriate framework for inference on this quantity. Let us therefore re-express (\ref{eq:standard_wvest}) as an M-estimator defined as follows:
\begin{equation}
	\hat{\nu}_{j}^2 := \underset{\nu_j^2 \in \mathbf{N}}{\argzero} \, \sum_{t=1}^{M_j} \psi(W_{j,t},\nu_j^2),
	\label{m.est}
\end{equation}
where $\mathbf{N} \subset \mathbb{R}^+$ and $\psi(\cdot)$ is a score function ($\psi$-function) which can be unbounded or bounded with respect to $(W_{j,t})$. For bounded $\psi$-functions, popular choices include Huber's function and Tukey's Biweight function \citep[see e.g.][]{HaRoRoSt:86}. Based on standard properties of M-estimators in dependent data settings \citep[see e.g.][]{kunsch1984infinitesimal}, the following proposition states the sufficient conditions under which the WV estimator has a bounded influence function (IF) and is therefore robust. 
\begin{prop}
\label{theo.IF}
Assuming that $(W_{j,t})$ is a strictly stationary process, the IF of the estimator of WV is bounded if $\psi(\cdot)$ is bounded.
\end{prop}
Given this intuitive result whose proof can be found in Appendix~\ref{app:wv.if}, we therefore intend to deliver an M-estimator of the form in (\ref{m.est}) whose $\psi$-function can be bounded and directly estimates the quantity of interest $\nu_{j}^2$. The proposed approach is based on ``Huber's Proposal 2'' which was presented in \cite{Hube:81} and was aimed at the estimation of the scale parameter of the residuals in the linear regression framework. Without loss of generality, we assume that $\mathbb{E}[W_{j,t}]~=~0$ and consequently use this proposal by defining $r_{j,t}~:=~W_{j,t}/\nu_{j}$ as the standardized wavelet coefficients, thereby defining the proposed estimator as:
	\begin{equation}
	\hat{\nu}_{j}^2 := \underset{\nu_j^2 \in \mathbf{N}}{\argzero} \, \left[ \frac{1}{M_{j}}\sum_{t=1}^{M_j} \omega^{2} \left(r_{j,t};\nu_{j}^2,c\right) \,r^{2}_{j,t} - a(\nu^2_{j},c) \right],
	\label{hub.est}
	\end{equation}
where $\omega(\cdot)$ represents the weight function implied by the chosen $\psi$-function and $a(\nu^2_{j},c)$ is a correction term to ensure Fisher consistency at the marginal distribution of the wavelet coefficients $(W_{j,t})$. Indeed, this correction term is defined as
$$a(\nu^2_{j},c) := \mathbb{E}[\omega^{2} \left(r_{j,t};\nu_{j}^2,c\right) \,r^{2}_{j,t}],$$
where the latter expectation is taken over the marginal distribution of $(W_{j,t})$ with variance $\nu_j$. It can be noticed that, if the tuning constant $c~\to~\infty$, we have that $\omega(\cdot)~\to~1$ and $a(\nu^2_{j},c)~\to~1$ thereby delivering the estimator in (\ref{eq:standard_wvest}) which implies that the tuning constant $c$ can be chosen to regulate the trade-off between robustness and efficiency of the resulting estimator. A discussion about the choice of the constant $c$ can be found in Appendix~\ref{choice_c} which highlights how this constant can be implicitly chosen by defining the level of statistical efficiency required with respect to the standard estimator. As mentioned, the term $a(\nu^2_{j},c)$ depends on the marginal distribution of the stationary process which is assumed for the wavelet coefficients $(W_{j,t})$ and on the specific weight function $\omega(\cdot)$. The exact analytic form of this term therefore may be complicated to derive when considering distributions other than the Gaussian or other symmetric distributions.

\begin{remark}
\label{remark-cons-correct}
In the case where the wavelet coefficients are assumed to come from a Gaussian distribution, the correction term $a(\nu^2_{j},c)$ can be expressed as $a_{\psi}(c)$ since it only depends on the value of the tuning constant $c$, and can be found explicitly using the results of \citet{dhrymes2005moments}. This is the case when the time series is itself Gaussian \citep[see e.g.][for a discussion]{Perc:16} or can be assumed as an approximation given the averaging nature of the wavelet filter. On the other hand, if the marginal distribution of the wavelet coefficients is non-Gaussian, the term $a(\nu^2_{j},c)$ could eventually be numerically approximated once (based on the standardized simulated values from the assumed distribution with known or estimated parameters) and used for all scales $\tau_j$ and would need to be accounted for in the subsequent inference phase.
\end{remark}

Having defined the M-estimator of WV in (\ref{hub.est}), let us now list a set of conditions which allow us to derive the asymptotic properties of this estimator. Firstly, letting $\nu_{j,0}^2$ represent the true WV at scale $\tau_j$, we define the following conditions for the $\psi$-function:
  \begin{enumerate}[label=\bfseries (C\arabic*), leftmargin = 1.5cm]
    \item The Bouligand derivative of $\psi(\cdot)$ is continuous almost everywhere. \label{cond.bdiff}
    \item $\mathbb{E}[\psi(W_{j,t}, \nu_{j}^2)] = 0$ if and only if $\nu_{j}^2 = \nu_{j,0}^2$. \label{cond.wv_ident}
 \end{enumerate}
The first condition is a technical requirement that allows us to perform an expansion in order to represent $\hat{\nu}_j^2$ in an explicit form (in addition to the Tukey Biweight function, for example, the Huber $\psi$-function also respects this condition as shown in Lemma~\ref{bdiff} in Appendix~\ref{app:asy.norm.wv}). On the other hand, Condition \ref{cond.wv_ident} requires the true WV to be identifiable through the chosen function $\psi(\cdot)$. This condition is verified when choosing the Huber and Tukey Biweight $\psi$-functions and assuming the wavelet coefficient process $(W_{j,t})$ is Gaussian as shown in Appendix~\ref{app:wv.ident}. 

Given these technical conditions on the properties of the function $\psi(\cdot)$, we can now study the process-related conditions for which we define the filtration $\mathcal{F}_t~:=~(\dots, \, \epsilon_{t-1}, \, \epsilon_t)$ where $\epsilon_t$ are \textit{i.i.d.} random variables. With this definition we can now deliver the first process-related condition.
  \begin{enumerate}[label=\bfseries (C\arabic*), leftmargin = 1.5cm, resume]
    \item $(W_{j,t})$ is a strictly stationary process and can be represented as 
    \begin{equation*}
        W_{j,t} = g_j(\mathcal{F}_t),
         \label{cond.ergostationary}
    \end{equation*}
 where $g_j(\cdot)$ is an $\mathbb{R}$-valued measurable function such that $W_{j,t}$ is well defined.
 \end{enumerate}
Although not necessarily expressed in this form, Condition \ref{cond.ergostationary} is commonly assumed when studying asymptotics within a time series setting and is respected by a very general class of time series models \citep[see][]{wu2005nonlinear}. In our case, this condition is necessary in order to apply the functional dependence measure defined in \cite{wu2011asymptotic}.

At this stage, we further denote $W_{j,t}^{\star}~=~g_j(\mathcal{F}_t^{\star})$ as being the coupled version of $W_{j,t}$, where $\mathcal{F}_t^{\star}~:=~(\dots, \, \epsilon_0^{\star}, \, \dots, \, \epsilon_{t-1}, \, \epsilon_t)$, with $\epsilon_0$ and $\epsilon_0^{\star}$ also being \textit{i.i.d.} random variables. Hence, the two processes $(W_{j,t})$ and $(W_{j,t}^{\star})$ depend on filtrations that only differ by one element, i.e. $\epsilon_0$ and $\epsilon_0^{\star}$, therefore implying that $W_{j,t}^{\star}~=~W_{j,t}$ for $t~<~0$. Based on this definition, we can define the functional dependence measure given in \citet{wu2011asymptotic}:
\begin{equation*}
    \delta_{t, q}^j := \Vert W_{j,t} - W_{j,t}^{\star}\Vert_q,
\end{equation*}
where $\Vert Z \Vert_q~:=~\left( \mathbb{E}[| Z |^q]\right)^{\nicefrac{1}{q}}$ for $q~>~0$. This dependence measure can be interpreted as the expected impact of the innovation $\epsilon_0^{\star}$ on the moments of $W_{j,t}^{\star}$ with respect to its ``original'' path given by $W_{j,t}$. Using this defintion, we provide the final process-related condition.
  \begin{enumerate}[label=\bfseries (C\arabic*), leftmargin = 1.5cm, resume]
    \item The process $(W_{j,t})$ is such that $\sum_{t=0}^{\infty} \delta_{t,4}^j < \infty$.
    \label{cond.weakdep}
 \end{enumerate}
This condition can be interpreted as a requirement for the expected difference between the fourth-order moments of the processes $(W_{j,t})$ and $(W_{j,t}^{\star})$ to go to zero as $t~\to~\infty$, implying that the innovation $\epsilon_0^{\star}$ has a limited impact in time on how much $W_{j,t}^{\star}$ deviates from $W_{j,t}$ and, hence, the process $(W_{j,t})$ is a ``stable'' process \citep[see][]{wu2011asymptotic}.

\begin{remark}
In the case where the chosen filter for the wavelet decomposition belongs to the Daubechies family, Conditions \ref{cond.ergostationary} and \ref{cond.weakdep} can be placed directly on the d$^{th}$-order difference of the process $(X_t)$, that we will denote as $\Delta_t$, instead of the process $(W_{j,t})$. The advantage of this consists in the fact that, if the length of the wavelet filter at the first level is such that $L_1~\geq~d$ (where $d$ is the required differencing such that $(\Delta_t)$ is stationary), then we only need $(\Delta_t)$ to respect these conditions to ensure that all levels of decomposition respect them too. Indeed, in the case of a Daubechies wavelet filter, all levels of wavelet coefficients simply correspond to a deterministic linear combination of $(\Delta_t)$ and we would therefore have $g_j(\cdot)~= \gamma_j g(\cdot)$ and $\delta_{t,q}^j~=~\lambda_j \delta_{t,q}$ where $g(\cdot)$ and $\delta_{t,q}$ would be uniquely related to $\Delta_t$ and constants $\gamma_j$ and $\lambda_j$ only depend on the chosen wavelet filter. For example, the process $(W_{j,t})$ in Conditions \ref{cond.ergostationary} and \ref{cond.weakdep} would be replaced by the process $(\Delta_t~=~X_t - X_{t-1})$ in the case of the Haar wavelet filter and hence we would have $\Delta_t~=~g(\mathcal{F}_t)$ and $\delta_{t,q}~=~\Vert \Delta_t - \Delta_t^{\star}\Vert_q$.
\end{remark}

We can now determine the asymptotic properties of the proposed M-estimator of WV in (\ref{hub.est}). For this reason, let us further define $\bm{W}_t~:=~[W_{j,t}]_{j = 1, \hdots, J}$ as the vector of wavelet coefficients at time $t$ and $\hat{\bm{\nu}}~:=~[\hat{\nu}_j^2]_{j = 1, \hdots, J}$ as the vector of estimated WV using the proposed M-estimator. Moreover, we define the projection operator \citep[see][]{wu2011asymptotic} as
$$\mathcal{P}_t \cdot := \mathbb{E}\left[\cdot | \mathcal{F}_t\right] - \mathbb{E}\left[\cdot | \mathcal{F}_{t-1}\right].$$
This operator therefore represents a measure of how much the conditional expectation of a process can change once the immediately previous information is removed. As for the previously defined functional dependence measure, intuitively the projection operator should not be too sensitive if the underlying process is stable \citep[in the sense of ][]{wu2011asymptotic}.

Using the above definitions, we finally define the quantities $\bm{D}_0~:=~\sum_{t=0}^{\infty} \mathcal{P}_0 \psi(\bm{W}_{t},\bm{\nu})$ and $\bm{M}~:=~\mathbb{E}[-\nicefrac{\partial}{\partial \bm{\nu}} \, \psi(\bm{W}_t, \bm{\nu})]$ to deliver the following theorem on the asymptotic distribution of the proposed estimator $\hat{\bm{\nu}}$.

\begin{thm}
\label{thm.asy.nuhat}
Under Conditions \ref{cond.bdiff} to \ref{cond.weakdep} and assuming that the function $\psi(\cdot)$ is bounded, we have that the estimator $\hat{\bm{\nu}}$ has the following asymptotic distribution
$$	\sqrt{T}\left(\hat{\bm{\nu}}-\bm{\nu}\right) \xrightarrow{\mathcal{D}} \mathcal{N}\left(\bm{0}, \bm{V}\right),$$
where $\bm{V} = \bm{M}\,\mathbb{E}[\bm{D}_0 \bm{D}_0^{\intercal}]\,\bm{M}^{\intercal}$.
\end{thm}
\noindent The proof of this theorem can be found in Appendix \ref{app:asy.norm.wv} where the proofs for Condition~\ref{cond.wv_ident} (for Huber and Tukey Biweight functions in the Gaussian setting) and consistency of $\hat{\nu}_j^2$ can also be found.

\begin{remark}
The asymptotic covariance matrix $\bm{V}$ is a long-run covariance matrix which can be estimated via different methods. For example, the moving block bootstrap, the batched-mean estimator \citep[see][]{zhang2017gaussian} or the progressive batched-mean method generalized from the idea in \cite{kim2013progressive} are available methods for such a purpose. In the setting of this paper, we generally assume a parametric family $F_{\bm{\theta}}$ for $(X_t)$ and, for this reason, the parametric bootstrap can also be considered.
\end{remark}

Aside from studying the behaviour of the proposed M-estimator as a first-step estimator for the RGMWM (Section~\ref{sec:simulations}), as an additional exercise in Appendix~\ref{sim_wv} we compare the behaviour of $\hat{\nu}_j^2$ with the standard WV estimator $\tilde{\nu}_j^2$ and the median-type robust estimator in \citet{mondal2012wavelet}. In the latter simulation it appears clearly that the proposed M-estimator is the best alternative to the standard estimator in the uncontaminated setting and the best overall in the contaminated setting. Based on the robustness properties of the proposed estimator $\hat{\bm{\nu}}$ (for bounded  $\psi$-functions) and its asymptotic properties, this new estimator provides a suitable tool to perform robust scale-based analysis of variance for time series \citep[see e.g.][and references therein]{percivalbook}. More importantly, it delivers a first-step estimator with adequate properties based on which it is possible to perform robust parametric inference for a wide range of time series models as well as for large data sets as discussed in the following section.

\section{Robust Generalized Method of Wavelet Moments}
\label{sec:rgmwm}

The properties of the proposed M-estimator of WV can be transferred directly within the GMWM framework \citep[see][]{gmwm}. Indeed, as suggested in \cite{GuMoVi:14}, one can replace the standard estimator used in the GMWM with a robust estimator which, in this case, is the proposed M-estimator allowing us to deliver the RGMWM defined as:
\begin{equation}
    \hat{\bm{\theta}} := \underset{\bm{\theta} \in \bm{\Theta}}{\text{argmin}}\, (\hat{\bm{\nu}} - \bm{\nu}(\bm{\theta}))^{\intercal}\bm{\Omega}(\hat{\bm{\nu}} - \bm{\nu}(\bm{\theta})),
    \label{eq.rgmwm}
\end{equation}
where $\bm{\Omega}$ is a symmetric positive definite weighting matrix chosen in a suitable manner (for example, one can choose $\bm{\Omega}~:=~\hat{\bm{V}}^{-1}$ where $\hat{\bm{V}}$ is a suitable estimator of $\bm{V}$, see \citealp{gmwm}). Moreover, the robustness (bounded IF) of the RGMWM estimator is inherited from the robustness of $\hat{\bm{\nu}}$ as shown in \citet{GeRo:03} in the indirect inference framework.

With this in mind, in the following paragraphs we list and discuss the conditions for the consistency and asymptotic normality of the RGMWM which summarize and reduce those in \citet{gmwm} for the standard GMWM. For this reason, let us define $\hat{\mathbf{\Omega}}$ as an estimator of $\bm{\Omega}$ and $||\cdot||_S$ as the matrix spectral norm which allow us to state the following conditions:

 \begin{enumerate}[label=\bfseries (C\arabic*), leftmargin = 1.5cm, resume]
    \item $\bm{\Theta}$ is compact. \label{compact.theta}
    \item $\bm{\nu}(\bm{\theta})$ is continuous and differentiable for all $\bm{\theta} \in \bm{\Theta}$. \label{continuous}
    \item  For $\bm{\theta}_1,\, \bm{\theta}_2 \in \bm{\Theta}$, 
    $\bm{\nu}(\bm{\theta}_1) = \bm{\nu}(\bm{\theta}_2)$ if and only if $\bm{\theta}_1 = \bm{\theta}_2$. \label{ident}
    \item $|| \hat{\mathbf{\Omega}} - \bm{\Omega}||_S \overset{p}{\to} 0$. \label{omega.cond}
 \end{enumerate}


Condition~\ref{compact.theta} is commonly assumed but could be replaced by imposing  
other technical constraints if deemed more appropriate with respect to the parametric setting of reference (as proposed, for example, in Theorem 2.7 of \citealp{newey1994large}). Condition~\ref{continuous} is easy to verify and is respected for most intrinsically stationary processes. However, Condition~\ref{ident} is an essential one which is often hard to verify. In this case, with respect to the Haar wavelet filter (which is one of the most commonly used wavelet filters), the discussion in \citet{gmwm} and the results in \citet{greenhall1998spectral} support the identifiability of a large class of (latent) time series models and are extended in Appendix~\ref{app_ident} as well as in \citet{identlatent} thereby relaxing Condition~\ref{ident}. Finally, Condition~\ref{omega.cond} addresses the choice of the weighting matrix $\bm{\Omega}$ and its corresponding estimator $\hat{\bm{\Omega}}$. In this perspective, the RGMWM is consistent for any matrix $\bm{\Omega}$ that is symmetric positive definite and, therefore, one needs to select an estimator $\hat{\bm{\Omega}}$ that converges to the chosen matrix $\bm{\Omega}$. A final condition which has not been stated is the consistency of the WV estimator $\hat{\bm{\nu}}$ which is implied by Conditions~\ref{cond.wv_ident} to \ref{cond.weakdep} as seen in Section~\ref{sec:wv}.

With these conditions we can now state the consistency of the RGMWM estimator $\hat{\bm{\theta}}$.

%
%
%
%

\begin{prop}
\label{th.cons.rgmwm}
	Under Conditions \ref{cond.wv_ident} to \ref{omega.cond} we have that $\hat{\bm{\theta}} \xrightarrow{\mathcal{P}} \bm{\theta}_0$.
\end{prop}

\noindent The proof of this proposition can be found in Appendix \ref{app_rgmwm_cons}. With this result, we can finally give the conditions for the asymptotic normality of $\hat{\bm{\theta}}$. For this reason, let us define
$$\bm{A}(\bm{\theta}_0) := \frac{\partial}{\partial \bm{\theta}^{\intercal}}\,\bm{\nu}(\bm{\theta})\Big|_{\bm{\theta} = \bm{\theta}_0},$$
which exists for a wide class of time series models \citep[see e.g.][]{Zhan:08}. Using this definition, we can state these final conditions.

\begin{enumerate}[label=\bfseries (C\arabic*), leftmargin = 1.5cm, resume]
    \item $\bm{\theta}_0 \in \text{Int}(\bm{\Theta})$. \label{interior}
        \item $\mathbf{H}(\bm{\theta}_0) := \bm{A}(\bm{\theta}_0)^{\intercal}\bm{\Omega} \bm{A}(\bm{\theta}_0)$ exists and is non-singular. \label{non.sing}
\end{enumerate}

Condition~\ref{interior} is a standard regularity condition while Condition~\ref{non.sing} is also usually assumed since it depends on the specific parametric model $F_{\bm{\theta}}$ from which the time series $(X_t)$ is generated and cannot therefore be verified in general. Since $\bm{\Omega}$ is non-singular by definition, we have that this condition relies mainly on the non-singularity of (the first $p$ rows of) $\mathbf{A}(\bm{\theta}_0)$ which is used, for example, to discuss Condition~\ref{ident} in Appendix~\ref{app_ident}. Finally, as for the results of consistency in Proposition~\ref{th.cons.rgmwm}, an additional condition for the asymptotic normality of the RGMWM is the asymptotic normality of $\hat{\bm{\nu}}$ which is stated in Theorem~\ref{thm.asy.nuhat}. Having discussed these conditions, we can use them to state the following lemma.

\begin{lemma}
\label{th.norm.rgmwm}
	Under Conditions \ref{cond.bdiff} to \ref{non.sing}, the estimator $\hat{\bm{\theta}}$ has the following asymptotic distribution
\begin{equation*}
	\sqrt{T}\left(\hat{\bm{\theta}}-\bm{\theta}_0\right) \xrightarrow[T\rightarrow\infty]{\mathcal{D}} \mathcal{N}\left(\mathbf{0},\mathbf{B} \mathbf{V} \mathbf{B}^{\intercal}\right),
\end{equation*}		
where $\mathbf{B} = \mathbf{H}(\bm{\theta}_0)^{-1} \mathbf{A}(\bm{\theta}_0)^{\intercal} \mathbf{\Omega}$.
\end{lemma}

The proof of Lemma \ref{th.norm.rgmwm} is provided in Appendix \ref{app_rgmwm_norm}. With the above results on the asymptotic properties of the RGMWM, the next paragraphs discuss some practical and theoretical advantages of the proposed robust framework.

\subsection{Discussion: Practical Properties and Extensions}
\label{Sec_practical-prop}

The RGMWM delivers various advantages that are mainly due to its two-step nature which allows it to benefit from the generality of the M-estimation framework presented in Section~\ref{sec:wv_mest}. A first advantage resides in the fact that the correction term needed to ensure Fisher consistency of the estimator defined in (\ref{hub.est}) only depends on the marginal distribution of the wavelet coefficients $(W_{j,t})$ which, if assumed to be Gaussian, can have an explicit form (for common $\psi$-functions) and only needs to be computed once for all levels $j$. A main advantage however resides in the fact that the dimension of the auxiliary parameter vector is always reasonable since in general $J~<~\log_2(T)$ which allows to make use of all the scales of WV without the need to select specific moments. This is not the case, for example, for Generalized Method of Moments (GMM) estimators where moment-selection is an important issue since, according to the model that is being estimated, the choice should fall on all moments (which can be highly impractical) or on moments that are more ``informative'' than others \citep[see e.g.][]{andrews1999consistent}. The RGMWM on the other hand can make use of all the possible scales of WV even for extremely large sample sizes, allowing it to preserve its statistical efficiency while gaining in terms of computational speed which is approximately of order $\mathcal{O}(T\log_2(T))$ while for the MLE, for example, it is roughly $\mathcal{O}(T^3)$. Appendix~\ref{sim_compeff} reports results on the computational time required to estimate the parameters of some (complex) models for sample sizes up to 10 million, confirming the considerable computational advantage of the RGMWM over both standard and robust alternatives. Moreover, the wavelet decomposition (and consequent variance estimation) is computationally efficient based on well-known algorithms \citep[see e.g.][]{rioul1992fast} and more recent approaches \citep[see e.g.][]{stocchi2018fast} allowing the RGMWM to be scalable. Nevertheless, a possible limitation of this approach is that it requires a large enough sample size to estimate more complex models (although, for example, it can already estimate four-parameter models with a sample size of $T~=~20$ if using a Haar wavelet filter).

For model comparison purposes, using its ``model-independent'' nature based on the robust weights of the proposed M-estimator in \eqref{hub.est}, the latter allows to graphically compare potential candidate models on the basis of the estimated WV as is routinely done, for example, with error characterization in the field of signal processing \citep[see e.g.][]{elsheimy08}. Indeed, decreasing linear trends in the log-log plot of the WV can indicate the presence of white noise or rounding-error models while increasing linear trends can indicate the presence of non-stationary components such as drifts and random walks whereas slight ``bumps'' in the plot can indicate the presence of ARMA components. In Section~\ref{sec:applications}, for example, the graphical display of the WV is used to detect and check the fit of the model when analysing a real data set on personal saving rates. Moreover, continuing with its model comparison advantages, the RGMWM estimator also delivers a general framework for robust goodness-of-fit tests and model selection. Indeed, the objective function in (\ref{eq.rgmwm}) can be used as a statistic for a goodness-of-fit test (Sargan-Hansen test or $J$-test) as proposed by  \citet{hansen1982large}, where the asymptotic distribution under the null hypothesis is chi-squared with $J-p$ degrees of freedom. Moreover, model selection criteria can also be built based on the (penalized) GMM objective function which in the RGMWM setting would be given by
\begin{equation}
    \label{GMM-objective}
   T(\hat{\bm{\nu}} - \bm{\nu}(\hat{\bm{\theta}}_k))^{\intercal}\hat{\bm{\Omega}}_k(\hat{\bm{\nu}} - \bm{\nu}(\hat{\bm{\theta}}_k)) + \Lambda\left(\hat{\bm{\theta}}_k,\hat{\bm{\Omega}}_k\right),
\end{equation}
where $\hat{\bm{\theta}}_k$ and $\hat{\bm{\Omega}}_k$ denote respectively the estimated parameter vector and weighting matrix for the $k^{th}$ model within a set of $K$ candidate models, while $\Lambda\left(\hat{\bm{\theta}}_k,\hat{\bm{\Omega}}_k\right)$ is a possible penalty term. Penalized objective functions have been proposed for model and moment selection for GMM estimators such as, for example, \citet{andrews1999consistent} and successively \citet{AnLu:01} who proposed penalty terms that reflect the number of moment conditions or \citet{zhang2015ModelSelection} who derived a penalty term using the covariance penalty criterion of \citet{efron2004estimation}. Nevertheless, since these model selection criteria only rely on a consistent estimator for the model's parameters $\bm{\theta}_k$, a robust version is (almost) readily available by using the RGMWM framework. Moreover, the estimator $\hat{\bm{\Omega}}_k$ in (\ref{GMM-objective}) can be made model-independent by choosing $\hat{\bm{\Omega}}_k := \hat{\bm{\Omega}}$ (e.g. based only on the empirical WV) making model comparison computationally more efficient. While the $J$-test statistic will be used in the analysis of real data in Section~\ref{sec:applications}, the study of a possible implementation of a robust model selection criterion  is left for further research.

Finally, aside from providing the basis for model-independent outlier-detection \citep[which can be of great importance for fault-detection algorithms, see for example][and references therein]{guerrier2012fault}, another advantage of the RGMWM estimator is that it can easily be extended to more complex settings such as multivariate time series \citep[see e.g.][]{mgmwm} or to random fields \citep[see e.g.][]{mondal2012wavelet, mondal2012mest} thereby delivering a computationally efficient framework for robust inference in these settings as well. 


\section{Simulation Studies}
\label{sec:simulations}

The aim of this section is to show that the RGMWM has a reasonable performance in settings where there is no data contamination and has a better performance than the classical (and possibly robust) alternatives when the data are contaminated. Concerning the robust alternatives, there is a lack of implemented and generally available (or usable) robust methods for parametric inference on time series models. For this reason, we were only able to successfully implement two robust estimators with which to compare the proposed RGMWM estimator: the Yule-Walker estimator based on the robust autocovariance estimator (YW) \citep[as used for example in][]{sarnaglia2010robust} and the indirect inference estimator based on the YW estimator \citep[as proposed in][]{GeRo:03} using a Tukey Biweight function with tuning constant $c = 2.2$ (chosen based on preliminary simulation studies in order to be highly robust). In the latter case, AR($p^*$) models were used as auxiliary models with $p^* = p + 1$ where, in this case, $p$ represents the number of parameters in the models of interest (excluding the innovation variance parameter). Since the YW estimator is appropriate for autoregressive models while the indirect inference estimator is used for all other models, we will denote both with a common acronym i.e. R-YW (for Robust Yule-Walker based estimators). On the other hand, the GMWM and RGMWM estimators are made available through the open-source \texttt{R} package \texttt{gmwm}\footnote{The \texttt{gmwm} package can be downloaded from \url{https://github.com/SMAC-Group/gmwm}.} where the default options are the Haar wavelet filter and a diagonal matrix for the weighting matrix $\hat{\bm{\Omega}}$ with elements proportional to the estimated variance of $\hat{\nu}_j^2$ (which was used for the simulations also to make more reasonable comparisons with the indirect inference estimators which were based on the identity matrix). The Tukey Biweight was used also for the RGMWM with tuning constant based on an asymptotic efficiency of 60\% thereby guaranteeing high robustness of the resulting estimator.

For the simulation studies different types of contamination were used, going from scale-contamination to additive and replacement outliers as well as patchy outliers and level-shifts. Innovation-type contamination was not considered since it does not affect the estimators much \citep[see][for an overview of different contamination settings]{MaMaYo:06, maronna2019robust}. We denote the proportion of contaminated observations with $\epsilon$ and the size of contamination (i.e. the variance of the observations which are added to the uncontaminated observations) with $\sigma_{\epsilon}^2$. Finally, when dealing with level-shifts, we denote $\mu_{\epsilon_i}$ as the size of the $i^{th}$ shift in level.

The performance of these estimators is investigated on the following models and contamination settings:
\begin{itemize}
	\item \textbf{AR(1)}: a zero-mean first-order autoregressive model with parameter vector $[\rho_1 \,\,  \upsilon^2]^{\intercal} = [0.9 \,\, 1]^{\intercal} $, scale-based contamination at level $j = 3$, $\epsilon = 0.01$ and $\sigma_{\epsilon}^2 = 100$;
	\item \textbf{AR(2)}: a zero-mean second-order autoregressive model with parameter vector $[\rho_1 \,\, \rho_2 \,\, \upsilon^2]^{\intercal} = [0.5 \,\, -0.3 \,\, 1]^{\intercal} $, isolated outliers, $\epsilon = 0.05$ and $\sigma_{\epsilon}^2 = 9$;
	\item \textbf{ARMA(1,2)}: a zero-mean autoregressive-moving average model with parameter vector $[\rho \,\, \varrho_1 \,\, \varrho_2 \,\, \upsilon^2]^{\intercal} = [0.5 \,\, -0.1 \,\, 0.5 \,\, 1]^{\intercal}$, and level-shift contamination with $\epsilon = 0.05$, $\mu_{\epsilon_1} = 5$ and $\mu_{\epsilon_2} = -3$;
	\item \textbf{ARMA(3,1)}: a zero-mean autoregressive-moving average model with parameter vector $[\rho_1 \,\, \rho_2 \,\, \rho_3 \,\, \varrho_1 \,\, \upsilon^2]^{\intercal} = [0.7\,\,0.3\,\,-0.2\,\,0.5\,\,2]^{\intercal}$, patchy outliers, $\epsilon = 0.01$ and $\sigma_{\epsilon}^2 = 100$;
	\item \textbf{SSM}: a state-space model $(X_t)$ interpreted as a composite (latent) process in certain engineering applications. This model is defined as
	\begin{equation*}
		\begin{aligned}
			Y_t^{(i)}=&\rho_{(i)} Y_{t-1}^{(i)} + W_t^{(i)}, 
			\;\;\; W_t^{(i)} \stackrel{\iid}{\sim} \mathcal{N}(0,\upsilon_{(i)}^{2}) \\ 
			X_t=&\sum_{i=1}^2 Y_t^{(i)} + Z_t,
			 \;\;\; Z_t \stackrel{\iid}{\sim} \mathcal{N}(0,\sigma^2)
		\end{aligned}
	\end{equation*}
	with parameter vector
	\begin{equation*}
		[\rho_{(1)} \,\, \upsilon_{(1)}^2 \,\,  \rho_{(2)} \,\, \upsilon_{(2)}^2 \,\, \sigma^2]^{\intercal} = [0.99\,\,0.1\,\,0.6\,\,2\,\,3]^{\intercal},
	\end{equation*}
	isolated outliers, $\epsilon = 0.05$ and $\sigma_{\epsilon}^2 = 9$.
\end{itemize}

To measure the statistical performance of the estimators we choose to use a robust and relative version of the Root Mean Squared Error (RMSE) defined as follows
	\begin{equation}
	\text{RMSE*} := \sqrt{\med\left(\frac{\hat{\theta}_i-\theta_{i,0}}{\theta_{i,0}}\right)^2 + \mad\left(\frac{\hat{\theta}_i}{\theta_{i,0}}\right)^2}\,\,, \nonumber
\end{equation}		
with $\med(\cdot)$ representing the median, $\mad(\cdot)$ the median absolute deviation and $\hat{\theta}_{i}$ and $\theta_{{i},0}$ representing the ${i}^{th}$ element of the estimated and true parameter vectors respectively. Finally, for each simulation, the number of simulated samples is 500 while the sample size is $T = 10^3$ which delivers $J = 9$ scales for the GMWM-type estimators. 

\begin{landscape}
\begin{figure}
\begin{center} 
\includegraphics[width=16cm]{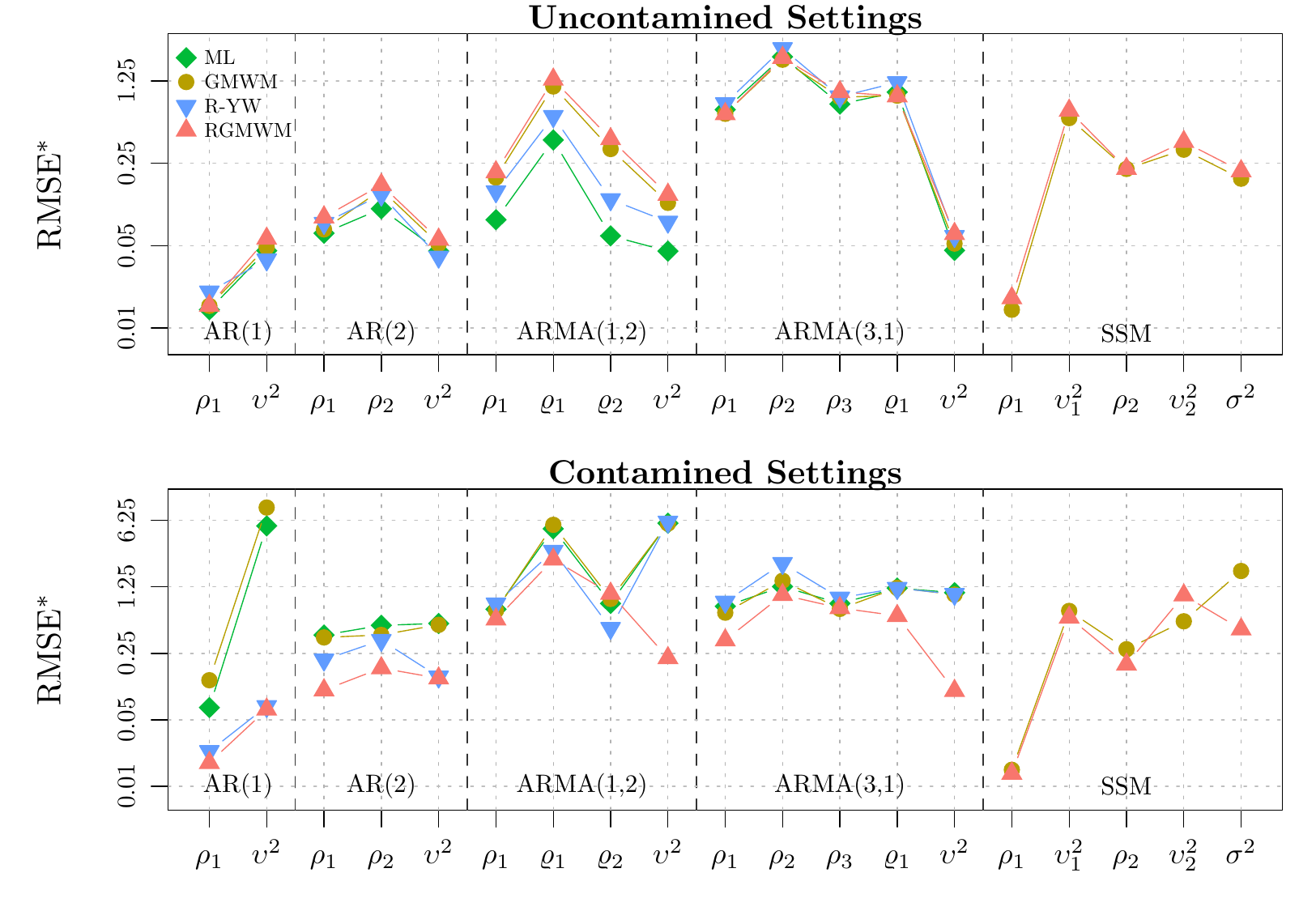}
 	\caption{Top row: logarithm of the RMSE* of the estimators in an uncontaminated setting. Bottom row: logarithm of the RMSE* of the estimators in a contaminated setting. R-YW represents the YW estimator for the \textbf{AR(1)} and \textbf{AR(2)} models while it represents the indirect inference estimator for the other models.}	
	\label{fig:ts.sim}
\end{center}
\end{figure}
\end{landscape}

Figure \ref{fig:ts.sim} displays the (logarithm of the) RMSE* of the estimators in both uncontaminated and contaminated settings for all the models presented above. When considering the \textbf{AR(1)} and \textbf{AR(2)} and \textbf{ARMA(3,1)} models, the RGMWM does not lose much in uncontaminated settings while it performs generally as well or better than the R-YW estimator in contaminated ones, especially concerning the variance parameter $\upsilon^2$ in the \textbf{ARMA(3,1)}. As for the \textbf{ARMA(1,2)} model, the RGMWM is not as efficient as the others in the uncontaminated case while it adequately bounds the influence of outliers in contaminated ones, performing generally better than the R-YW estimator, especially for the variance parameter $\upsilon^2$ as in the case of the \textbf{ARMA(3,1)} model. For the \textbf{SSM} model, standard estimators in their default implementation did not appear to be numerically stable while, to the best of our knowledge, robust alternatives have never been implemented. From Figure~\ref{fig:ts.sim}, it can be seen how the RGMWM is extremely close to the GMWM in uncontaminated settings while it remains more stable than the GMWM in the contaminated ones. When considered jointly with the additional simulation study in Appendix~\ref{sim_compeff}, this simulation exercise shows that the RGMWM provides a computationally efficient and numerically stable method to robustly estimate the parameters of many linear state-space models which, to date, has been almost unfeasible in practice with alternative robust estimators. Indeed, Appendix~\ref{sim_compeff} illustrates the computational efficiency of the RGMWM for time series models with a moderately high number of parameters for sample sizes of $T = 10^7$ for which the RGMWM is computed in just over a minute, denoting the added value of this approach since these models and sample sizes are extremely common, for example, in the natural sciences and engineering. As a final note, since this simulation study does not use a full weighting matrix  for $\hat{\bm{\Omega}}$, the efficiency of the (R)GMWM could be improved by choosing an alternative matrix.

\section{Application: Personal Saving Rates}
\label{sec:applications}

Having highlighted the properties of the RGMWM in a controlled simulated setting, in this section we conclude this work by presenting the results when using the RGMWM for an analysis on real data concerning personal saving rates. In addition, in Appendix~\ref{iner.sens} we present the results of an analysis on the measurement error issued from an inertial sensor based on a calibration sample of size $T=9\cdot 10^5$ that requires the estimation of a state-space model with 6 parameters. Indeed, the wide class of intrinsically stationary models for which the RGMWM can be used allows it to be employed in a large variety of applications where outliers and other types of data contamination can often occur. As mentioned above, in this section the RGMWM will be used to analyse data consisting in the monthly seasonally adjusted Personal Saving Rates (PSR) from January 1959 to  May 2015 provided by the Federal Reserve Bank of St. Louis\footnote{U.S. Bureau of Economic Analysis, Personal Saving Rate [PSAVERT], retrieved from FRED, Federal Reserve Bank of St. Louis; \url{https://fred.stlouisfed.org/series/PSAVERT}}. The study of PSR is an essential part of the overall investigation on the health of national and international economies since, within more general economic models, PSR can greatly impact the funds available for investment which in turn determine the productive capacity of an economy. Understanding the behaviour of PSR is therefore an important step in correct economic policy decision making. In this sense, \citet{slacalek2012drives} study the factors behind saving rates and investigate different models which, among others, are compared to the random-walk-plus-noise (local level) model (RWN). As opposed to the latter model, various time-varying models are proposed in the literature to explain precautionary PSR together with risk aversion in the light of different factors such as financial shocks or others \citep[see, for example,][]{videras2004behavioral,brunnermeier2008wealth}. Nevertheless, as emphasized  in \citet{pankratz2012forecasting}, modelling the time series with a stationary model, or a d$^{th}$-order non-stationary model such as an ARIMA, can be useful under many aspects such as, for example, to understand if a dynamic model is needed for forecasting and, if so, what kind of model is appropriate.

\begin{figure}[!ht]
\begin{center}
\includegraphics[width=11.5cm]{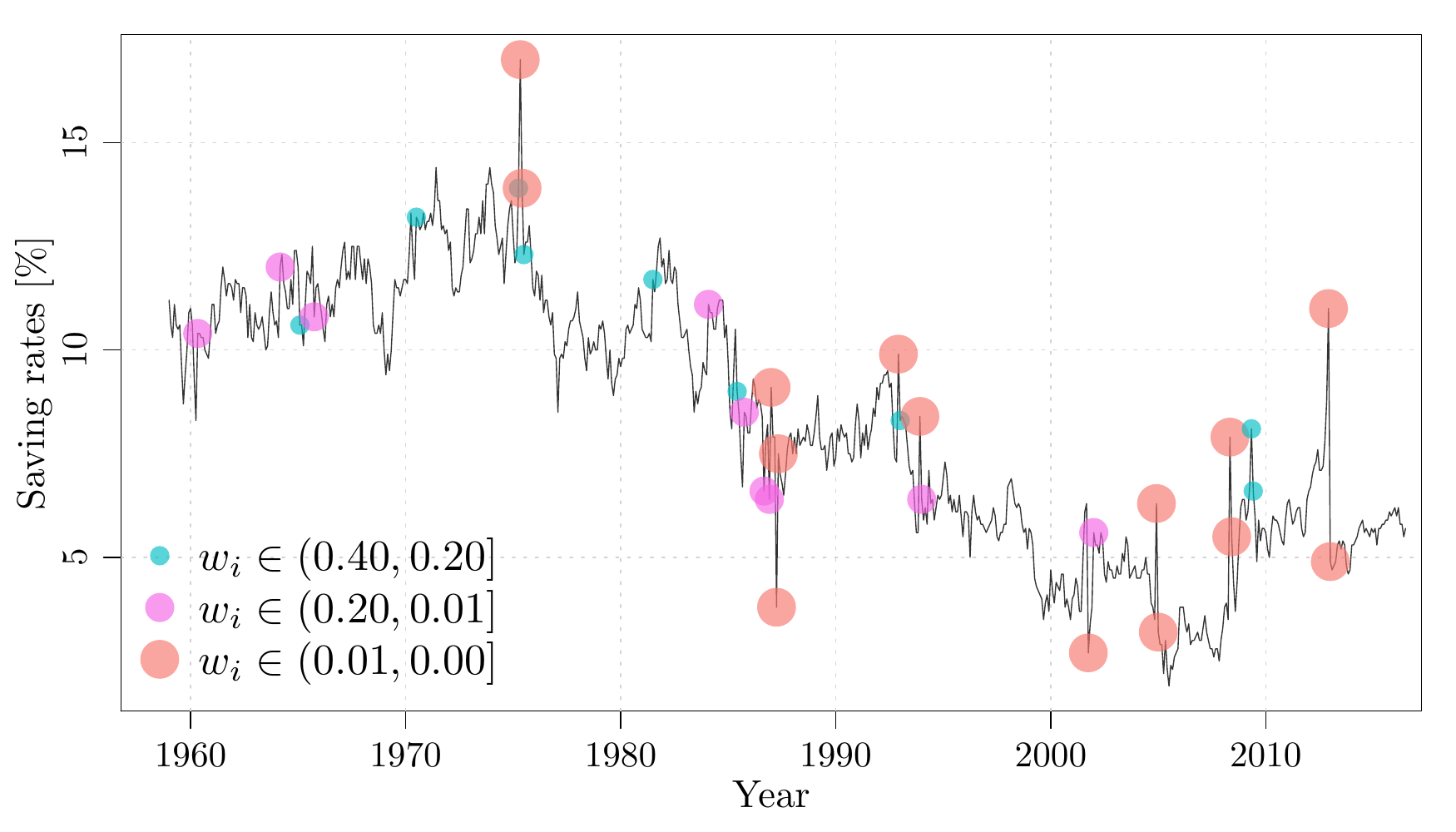}
 \includegraphics[width=11cm]{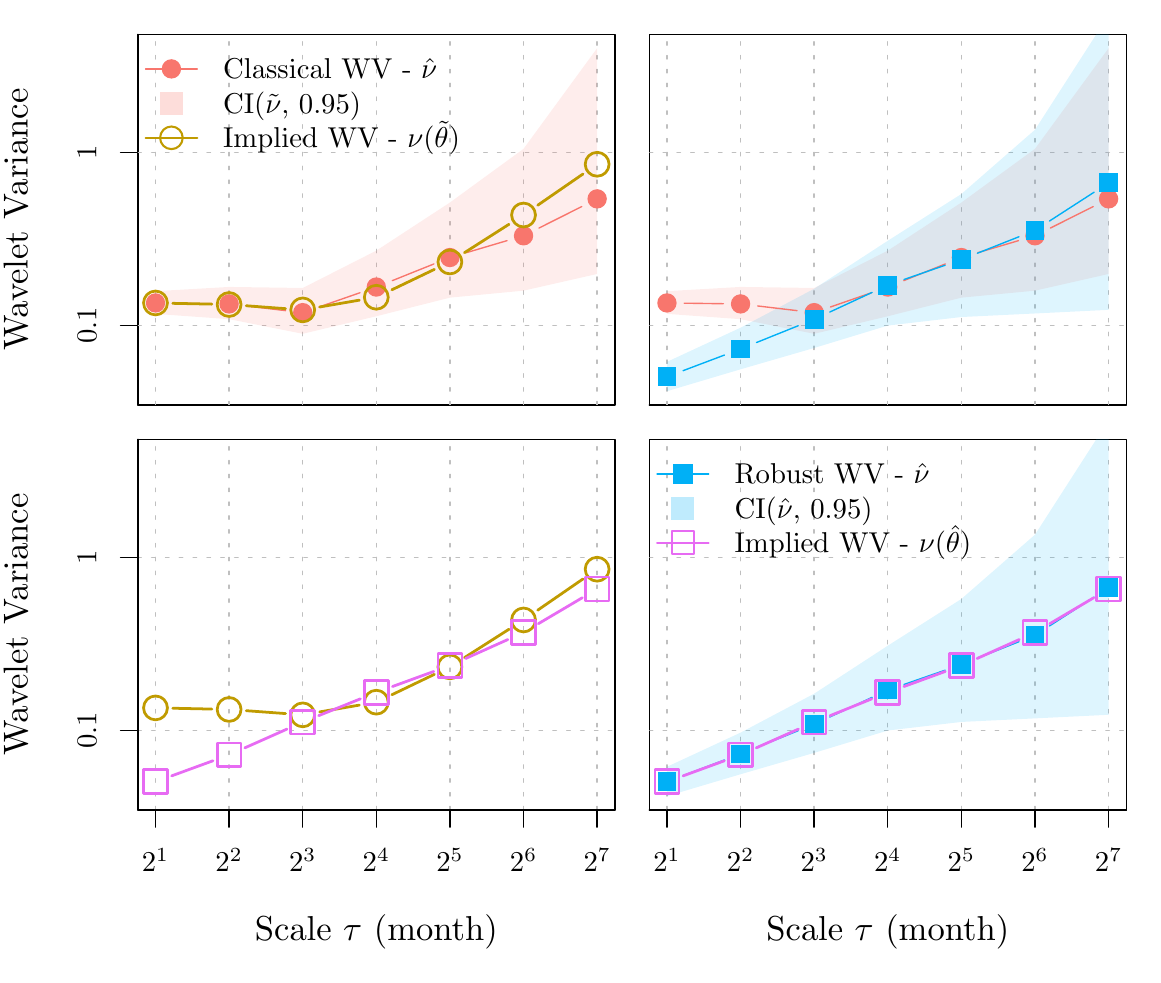}
 	\caption{Top figure: Saving rates time series with different types of points indicating outliers identified through the weights of the proposed M-estimator. Bottom figure: log-log scale WV plots for saving rates series with classic estimated WV superposed with model-implied WV based on the parameters estimated through the GMWM (top left); classic and robust estimated WV with respective confidence intervals superposed (top right); classic and robust model-implied WV based on the GMWM and RGMWM estimates respectively (bottom left); robust estimated WV superposed with model-implied WV based on the parameters estimated through the RGMWM estimator (bottom right).}
	\label{fig:savings}
\end{center}
\end{figure}

In this example, we consider the RWN model and we use the WV log-log plot and a $J$-test (see Section~\ref{Sec_practical-prop}) to understand what kind of model could fit the time series. Considering both the visual fit of the WV of the candidate models in the plots as well as the p-values of the $J$-tests (for which the null hypothesis is that the model fits the data well), we find that a random walk plus an ARMA(2,1) appears to be a good fit and therefore, in this case, we have that the ``noise'' in the RWN model is an ARMA(2,1) ($J$-test p-values are 0.109 and 0.279 for GMWM and RGMWM respectively). This can be seen in Figure~\ref{fig:savings} where, in the top part, the saving rate time series is represented along with the identified outliers and, in the bottom part, we see the log-log representation of the classic and robust estimated and model-implied WV respectively. Indeed, for the bottom part, the diagonal plots show the classic and robust estimations respectively, each with the estimated WV and the WV implied by the estimated RWN model; when the difference in the two lines lies within their confidence intervals, the chosen model can be considered as adequate. The off-diagonal plots compare the classic and robust estimated WV (upper off-diagonal) and the WV implied by the GMWM and RGMWM model parameter estimates (lower off-diagonal). 

It can be seen how there is a significant difference between the standard and robust WV estimates, especially at the first scales where the confidence intervals of the estimated WV do not overlap (upper off-diagonal plot). This leads to a difference in the model-implied WV whose parameters have been estimated through the GMWM and RGMWM (lower off-diagonal plot).

The estimated parameters of the RWN model using the GMWM and RGMWM are given in Table \ref{tab:savings} along with their respective confidence intervals. 
	\begin{table}
		\caption{Random Walk plus ARMA(2,1) model estimates for the PSR data. Estimated parameters with GMWM and RGMWM estimators with $\gamma^2$ being the random walk parameter, $\rho_i$ the $i^{th}$ autoregressive parameter, $\varrho$ the moving average parameter and $\sigma^2$ the innovation variance of the ARMA(2,1) model. Confidence intervals (CI) based on the approach used in \citet{gmwm}.}
		\centering
		\footnotesize
		 \begin{tabular}{ccccc}  
		    \toprule
		    & \multicolumn{2}{c}{GMWM} & \multicolumn{2}{c}{RGMWM} \\
		    \cmidrule(r){2-3} \cmidrule(r){4-5}
		        & Estimate & CI($\cdot$, 95\%) & Estimate & CI($\cdot$, 95\%) \\
		    \midrule
			$\gamma^2$ & $7.95\cdot 10^{-2}$ & $ (\phantom{-}3.67\cdot 10^{-2} \, ; \, 1.11\cdot 10^{-1}) $ & $5.85\cdot 10^{-2}$ & $ (1.54\cdot 10^{-2} \, ; \, 9.97\cdot 10^{-2})$ \\
			$\rho_1$ & $1.64\cdot 10^{-1}$ & $ (\phantom{-}5.93\cdot 10^{-2} \, ; \, 2.89\cdot 10^{-1}) $ & $6.00\cdot 10^{-1}$ & $ (4.48\cdot 10^{-1} \, ; \, 7.55\cdot 10^{-1})$ \\
			$\rho_2$ & $3.06\cdot 10^{-3}$ & $ (-1.31\cdot 10^{-1} \, ; \, 1.48\cdot 10^{-1}) $ & $1.84\cdot 10^{-1}$ & $ (3.10\cdot 10^{-2} \, ; \, 2.46\cdot 10^{-1})$ \\
			$\varrho$ & $2.43\cdot 10^{-1}$ & $ (\phantom{-}2.02\cdot 10^{-1} \, ; \, 2.81\cdot 10^{-1}) $ & $2.92\cdot 10^{-1}$ & $ (2.28\cdot 10^{-1} \, ; \, 3.45\cdot 10^{-1})$ \\
			$\sigma^2$ & $3.14\cdot 10^{-1}$ & $ (\phantom{-}2.59\cdot 10^{-1} \, ; \, 3.85\cdot 10^{-1}) $ & $1.32\cdot 10^{-1}$ & $ (8.59\cdot 10^{-2} \, ; \, 1.80\cdot 10^{-1})$ \\
		    \bottomrule
		  \end{tabular}
		  \label{tab:savings}
	\end{table}
There are two main differences between the two estimations: (i) the estimates of the first autoregressive parameter $\rho_1$ and innovation variance $\sigma^2$ are significantly different; (ii) the second autoregressive parameter $\rho_2$ is not significant using the GMWM. These differences highlight how the conclusions concerning parameter values and model selection can considerably change when outliers are present in the data. Indeed, the choice of the model would then affect the decisions taken towards the selection of appropriate causal and dynamic models to better explain the behaviour of saving rates. The selected model based on the robust fit can in fact be interpreted as a sum of latent models along the lines given in \citet{slacalek2012drives} where the ARMA(2,1) can be seen as a sum of two AR(1) models \citep[see][]{granger1976time} where each of them represents, for example, the reaction of PSR to changes in uncertainty (affected by unemployment) and interest rates, respectively, while the random walk describes the continuous fluctuations of target wealth which also drives PSR. 

The additional benefit of the RGMWM, and more specifically of the proposed M-estimator of WV, is also to deliver weights that allow to identify outliers which may not be visible simply by looking at the time series. As shown in the top part of Figure \ref{fig:savings}, the outliers identified by the RGMWM can be interpreted in the light of the national and global economic and political events. Limiting ourselves to the major identified outliers, the first one corresponds to a rise in the precautionary savings in the aftermath of the OPEC oil crisis and the 1974 stock market crash. In the months following October 1987 we can see an instability in the PSR with a rise and sudden fall linked to the ``Black Monday'' stock market crash which added to the savings and loans crisis which lasted to the early 1990s. This period also saw an economic recession where a rise in the saving rates, highlighted by the presence of high outliers, led to a drop in aggregate demand and bankruptcies. Finally, the various financial crises of the 21$^{st}$ century led to sudden and isolated rises in PSR as indicated again by the outliers.

\if0\blind
{
\section*{Acknowledgements}
We would like to thank Stephen Portnoy and Roger Koenker for their valuable inputs and suggestions throughout the development of this work. We would also like to thank Fabio Trojani for his interesting and helpful considerations. Moreover, we would like to thank Donald Percival and Debashis Mondal for sharing wavelet variance code and some data for the case studies presented in the paper and in its supplemental materials. Moreover, we thank Charles Greenhall for his precious insight into the discussion on the injective relationship between the spectral density and the wavelet variance. Finally, we would like to thank Mucyo Karemera, Marc-Olivier Boldi, Samuel Orso and Marco Avella for discussing the results of this work with us.
} \fi

\newpage

\bibliographystyle{plainnat}
\bibliography{references} 

\AtEndDocument{

\section*{Supplementary material}

# Appendices:
# Robust Two-Step Wavelet-Based Inference for Time Series Models

## Norm Notation

Given the use of different types of norm considered in these appendices, as a reference below we provide their notations and definitions:

- $|\boldsymbol{a}|_q := \left(\sum_{j=1}^{p} |a_j|^q\right)^{1/q}$ represents the $L_q$-norm where $\boldsymbol{a} = (a_1, \ldots, a_p)^\intercal \in \mathbb{R}^p$ (where $|\cdot|$ simply represents the $L_1$-norm).

- $\|X\|_q := (\mathbb{E}[|Z|^q])^{1/q}$ .

- $\|\boldsymbol{X}\|_S$ denotes the spectral norm of a matrix $\boldsymbol{X}$.

## A  Short literature review

A detailed discussion on robust estimation and inference methods for time series models can be found in Maronna et al. (2006), Chapter 8. An important part of the literature in this domain has dealt with time series models such as autoregressive and/or moving average models. For example, Kunsch (1984) proposes optimal robust M-estimators of the parameters of autoregressive processes by studying the properties of their influence function (see also Martin and Yohai, 1986). Denby and Martin (1979) develop a generalized M-estimator for the parameter of a first-order autoregressive process whereas Bustos and Yohai (1986), Allende and Heiler (1992) and de Luna and Genton (2001); Genton and Ronchetti (2003) extend the research to include moving average models using generalized M-estimation theory and/or indirect inference (see e.g. Gourieroux et al., 1993). Bianco et al. (1996) propose a class of robust estimators for regression models with ARIMA errors based on $\tau$-estimators of scale (Yohai and Zamar, 1988). Ronchetti and Trojani (2001) develop a robust version of the generalized method of moments (proposed by Hansen, 1982) for estimating the parameters of time series models in economics, while Ortelli and

1

Trojani (2005) further develop a robust efficient method of moments and Cizek (2016) propose a generalized method of trimmed moments. Mancini et al. (2005) develop optimal bias-robust estimators for a class of conditional location and scale time series models while La Vecchia and Trojani (2010) develop conditionally unbiased optimal robust estimators for general diffusion processes, for which approximation methods for computing integrals are proposed. Cizek (2008) studies the properties of a two-step least weighted squares robust time-series regression estimator and Agostinelli and Bisaglia (2010) propose a weighted maximum likelihood estimator for ARFIMA processes.

Two-step robust approaches can be built upon robust (Kalman) filtering or robust moment estimation. Robust estimators of moments, such as autocovariances, include Ma and Genton (2000); Lévy-Leduc et al. (2011); Chang and Politis (2016) (see also Rousseeuw and Croux, 1993), and for a review, see e.g. Dürre et al. (2015). They have been used by e.g. Molinares et al. (2009) as plugin estimators for ARFIMA models (see also Reisen and Molinares, 2012), by Sarnaglia et al. (2010) for the parameters of the periodic AR model with the Yule–Walker equation and by Bahamonde and Veiga (2016) for the GARCH(1,1). The idea of making the Kalman filter robust was originated with Masreliez and Martin (1977) and Cipra (1992) who propose robust modifications of exponential smoothing (see also Cipra and Hanzak, 2011 and Croux et al., 2010 for a multivariate version). For a robust version of the Holt-Winters smoother, see Gelper et al. (2010), and other proposals can be found in e.g. Ruckdeschel et al. (2014) and Calvet et al. (2015). Muler et al. (2009) develop a class of robust estimates for ARMA models that are closely related to robust filtering. Robustness properties of wavelet filtering have been studied for the identically and independently distributed (*i.i.d.*) case by Renaud (2002). Several robust local filters have been proposed so far since the median filter proposal from Tukey (1977): Bruce et al. (1994) pre-process the estimation of the wavelet coefficients via a "fast and robust smooth/cleaner"; Krim and Schick (1999) derive a robust estimator of the wavelet coefficients based on minimax description length; Härdle and Gasser (1984) develop a locally weighted smoothing using M-estimation and Fried et al. (2007) propose a non-parametric, weighted repeated median filter. Sardy et al. (2001) propose a robust wavelet-based estimator using a robust loss-penalized function, for which appropriately choosing the smoothing parameter is an important robustness issue as revealed, for example, by Cantoni and Ronchetti (2001).

# B Robust Wavelet Variance Properties

## B.1 Proof of Proposition 2.1

*Proof.* Let us define $\varepsilon$ as the contamination level and let $(X_t)$ be a process generated by a model in a neighborhood of $F_{\boldsymbol{\theta}}$, i.e. $F_\varepsilon = (1-\varepsilon)F_{\boldsymbol{\theta}} + \varepsilon \Delta_{\boldsymbol{z}}$, with small $\varepsilon > 0$. $F_\varepsilon$ is the standard contamination model with $\Delta_{\boldsymbol{z}}$ the Dirac function at $\boldsymbol{z} = X_t^\varepsilon$. Under the assumption of Proposition 2.1 and using the time series adaptation of the IF from Kunsch (1984), the IF of $\hat{\nu}_j^2$ is obtained by taking the Gâteaux derivative (of the estimating function seen as a functional of $F_\varepsilon$) with respect to $\varepsilon$ when $\varepsilon \to 0^+$, i.e.

$$\frac{\partial}{\partial \varepsilon}\left[(1-\varepsilon)\mathbb{E}_{F_{\boldsymbol{\theta}}}\left[\psi\left(W_{j,t}(F_\varepsilon), \hat{\nu}_j^2(F_\varepsilon)\right)\right]\right]_{\varepsilon \downarrow 0} + \frac{\partial}{\partial \varepsilon}\left[\varepsilon \psi\left(W_{j,t}(F_\varepsilon), \hat{\nu}_j^2(F_\varepsilon)\right)\right]_{\varepsilon \downarrow 0}$$

$$= -\mathbb{E}_{F_{\boldsymbol{\theta}}}\left[\psi\left(W_{j,t}(F_{\boldsymbol{\theta}}), \nu_j^2\right)\right] +$$

$$\mathbb{E}_{F_{\boldsymbol{\theta}}}\left[\frac{\partial}{\partial W_{j,t}} \psi\left(W_{j,t}, \nu_j^2\right)\right]_{W_{j,t}=W_{j,t}(F_{\boldsymbol{\theta}})} \frac{\partial}{\partial \varepsilon}\left[W_{j,t}(F_\varepsilon)\right]_{\varepsilon \downarrow 0} +$$

$$\mathbb{E}_{F_{\boldsymbol{\theta}}}\left[\frac{\partial}{\partial \nu_j^2} \psi\left(W_{j,t}(F_{\boldsymbol{\theta}}), \nu_j^2\right)\right] \frac{\partial}{\partial \varepsilon}\left[\hat{\nu}_j^2(F_\varepsilon)\right]_{\varepsilon \downarrow 0} + \psi\left(W_{j,t}(\boldsymbol{z}), \nu_j^2\right)$$

$$= -\mathbb{E}_{F_{\boldsymbol{\theta}}}\left[\psi\left(W_{j,t}(F_{\boldsymbol{\theta}}), \nu_j^2\right)\right] +$$

$$\mathbb{E}_{F_{\boldsymbol{\theta}}}\left[\frac{\partial}{\partial W_{j,t}} \psi\left(W_{j,t}, \nu_j^2\right)\right]_{W_{j,t}=W_{j,t}(F_{\boldsymbol{\theta}})} \text{IF}(\boldsymbol{z}, W_{j,t}, F_{\boldsymbol{\theta}}) +$$

$$\mathbb{E}_{F_{\boldsymbol{\theta}}}\left[\frac{\partial}{\partial \nu_j^2} \psi\left(W_{j,t}(F_{\boldsymbol{\theta}}), \nu_j^2\right)\right] \text{IF}(\boldsymbol{z}, \hat{\nu}_j^2, F_{\boldsymbol{\theta}}) + \psi\left(W_{j,t}(\boldsymbol{z}, F_{\boldsymbol{\theta}}), \nu_j^2\right) := \boldsymbol{K}$$

By the chain rule we have that $\text{IF}(\boldsymbol{z}, \hat{\nu}_j^2, F_{\boldsymbol{\theta}}) \propto \text{IF}(\boldsymbol{z}, W_{j,t}, F_{\boldsymbol{\theta}})$, therefore

$$\boldsymbol{K} \propto -\mathbb{E}_{F_{\boldsymbol{\theta}}}\left[\psi\left(W_{j,t}(F_{\boldsymbol{\theta}}), \nu_j^2\right)\right] +$$

$$\mathbb{E}_{F_{\boldsymbol{\theta}}}\left[\frac{\partial}{\partial W_{j,t}} \psi\left(W_{j,t}, \nu_j^2\right)\right]_{W_{j,t}=W_{j,t}(F_{\boldsymbol{\theta}})} \text{IF}(\boldsymbol{z}, \hat{\nu}_j^2, F_{\boldsymbol{\theta}}) +$$

$$\mathbb{E}_{F_{\boldsymbol{\theta}}}\left[\frac{\partial}{\partial \nu_j^2} \psi\left(W_{j,t}(F_{\boldsymbol{\theta}}), \nu_j^2\right)\right] \text{IF}(\boldsymbol{z}, \hat{\nu}_j^2, F_{\boldsymbol{\theta}}) + \psi\left(W_{j,t}(\boldsymbol{z}, F_{\boldsymbol{\theta}}), \nu_j^2\right) = 0.$$

Given that $\mathbb{E}_{F_{\boldsymbol{\theta}}}[\psi(W_{j,t}(F_{\boldsymbol{\theta}}), \nu_j^2)] = 0$ we finally obtain

$$\text{IF}(\boldsymbol{z}, \hat{\nu}_j^2, F_{\boldsymbol{\theta}}) \propto -\mathbf{D}^{-1}\psi\left(W_{j,t}(\boldsymbol{z}), \boldsymbol{\theta}\right)$$

with

$$\mathbf{D} = \mathbb{E}_{F_{\boldsymbol{\theta}}}\left[\frac{\partial}{\partial W_{j,t}}\psi\left(W_{j,t}, \nu_j^2\right)\right]_{W_{j,t}=W_{j,t}(F_{\boldsymbol{\theta}})} + \mathbb{E}_{F_{\boldsymbol{\theta}}}\left[\frac{\partial}{\partial \nu_j^2}\psi\left(W_{j,t}(F_{\boldsymbol{\theta}}), \nu_j^2\right)\right].$$

Since $\mathbf{D}$ does not depend on the contamination mass $\boldsymbol{z}$, the IF of the estimator of the WV is bounded if $\psi(\cdot)$ is bounded, thus concluding the proof. □

## B.2 Choice of Tuning Constant

Given that Theorem 2.1 provides an expression for the variance of the WV estimator, here we provide a brief discussion on the choice of the tuning constant $c$. The definition of this value is based on the desired level of efficiency compared to the classical estimator and varies according to the chosen $\psi$-function. However, an explicit and intuitive rule for the choice of this constant is available only when considering the process $(W_{j,t})$ as Gaussian. In the latter case, the estimator in (2.3) is the result of a minimization under the standard Gaussian assumption (i.e. zero mean and unit variance) and because of this we can obtain expressions for the variance of both the classical estimator and the robust estimator, which we denote as $\sigma_j^2$ and $\tilde{\sigma}_j^2(c)$ respectively. In this setting we see that these expressions depend solely on $c$ and therefore, for a general scale $\tau_j$ and defining $\alpha \in [0, 1]$ as the desired level of efficiency, a rule to select the tuning constant $c$, given a specific $\psi$-function, is to find the solution in $c$ to the expression

$$\frac{\sigma_j^2}{\tilde{\sigma}_j^2(c)} - \alpha = 0.$$

For example, choosing $\alpha = 0.95$ delivers a tuning constant $c \approx 7.88$ when using the Tukey $\psi$-function and $c \approx 2.38$ when using the Huber $\psi$-function (respectively $c \approx 4.97$ and $c \approx 1.22$ for $\alpha = 0.6$). The choice of the efficiency level is subjective and can be supported by a sensitivity analysis comparing the classical and the robust estimates starting from a low efficiency level (e.g. 0.5).

## B.3 Wavelet Variance Identifiability

In this appendix we discuss the identifiability of the WV when using the Huber and Tukey biweight $\psi$-functions. We first define the Tukey biweight function with redescending weights (see Beaton and Tukey, 1974). The biweight $\psi$-function delivers the following weights $\omega(\cdot)$

$$\omega_{[Bi]}(r_{j,t}; \nu_j^2, c) = \begin{cases} \left(\left(\frac{r_{j,t}}{c}\right)^2 - 1\right)^2 & \text{if } |r_{j,t}| \leq c \\ 0 & \text{if } |r_{j,t}| > c \end{cases}$$

and, if one supposes the normality for the wavelet coefficients, then the correction term $a(\nu_j^2)$ is

$$\begin{aligned} a_{[Bi]}(c) &= \mathbb{E}_\Phi \left[ \omega_{[Bi]}^2(r_{j,t}; \nu_j^2, c) r_{j,t}^2 \right] \\ &= \frac{1}{c^8} \mu_c^{10} - \frac{4}{c^6} \mu_c^8 + \frac{6}{c^4} \mu_c^6 - \frac{4}{c^2} \mu_c^4 + \mu_c^2 \end{aligned} \tag{B.1}$$

with $\mu_c^i$ being the $i$-th truncated moment under the standard normal distribution between $-c$ and $c$.

On the other hand, Huber's $\psi$-function has well known properties and has easily tractable derivatives when developing its asymptotic properties. Its weights are given by

$$\omega_{[Hub]}(r_{j,t}; \nu_j^2, c) = \min\left(1; \frac{c}{r_{j,t}}\right). \tag{B.2}$$

When using these functions, it is necessary to understand if they deliver functions which enable to identify the unknown parameter and state that

$$\mathbb{E}\left[\psi\left(W_{j,t}, \kappa^2\right)\right] = 0$$

if and only if $\kappa^2 = \nu_{0,j}^2$ (i.e. there is a unique solution for $\nu_j^2$). In this setting, we assume that the wavelet coefficient process $(W_{j,t})$ is a Gaussian process with zero mean. The latter assumption can always be verified regarding the mean constraint since, without loss of generality, all stationary models deliver zero-mean wavelet coefficients $(W_{j,t})$ with finite WV and many non-stationary models with stationary backward differences can also respect this condition by recentering the wavelet coefficients. This is the case for all stationary ARMA and various state-space models. However, the assumption of a Gaussian model for $(W_{j,t})$ issued from the previously mentioned models is a relatively strong one but (apart from the case where $(X_t)$ is itself Gaussian) it is a frequently assumed condition for the wavelet coefficients and, according to the type of process, could be a reasonable approximation due

to the averaging nature of the filter.

Having underlined this, let us start with the identifiability of the Huber weight function and, for this, let

$$Z = \begin{cases} r_{j,t}^2 & \text{for } |r_{j,t}| \leq c \\ c^2 & \text{for } |r_{j,t}| > c \end{cases}$$

with $r_{j,t} = W_{j,t}/\kappa$, $\kappa^2 \in \mathbb{R}^+$ and let us consider the function $\mathbb{E}[\psi(W_{j,t}, \kappa^2)]$. For Huber weights we define $q(r_{j,t}, c) := \mathbb{E}[Z - a_\psi(c)]$ where $a_\psi(c)$ is a constant for a given $c$. For identifiability we need to prove that $q(r_{j,t}, c)$ has a unique solution in $\nu_{0,j}^2$ and to do so we prove that its derivative is a strictly monotone function in $\kappa^2$. Indeed, we have by definition that $\mathbb{E}[\psi(W_{j,t}, \kappa^2)] = 0$ if $\kappa^2 = \nu_{0,j}^2$ and if the derivative of $q(r_{j,t}, c)$ is strictly monotone then the solution is unique. Let us denote $\mathbb{P}[A]$ as the probability of event $A$, $\alpha := r_{j,t}\frac{\kappa}{\nu_j}$ and $\gamma := c\frac{\kappa}{\nu_j} > 0$, then we have that

$$\begin{aligned} \mathbb{E}[Z - a_\psi(c)] &= \mathbb{E}[Z] - a_\psi(c) \\ &= \mathbb{E}\left[Z \Big| |r_{j,t}| \leq c\right] \mathbb{P}[|r_{j,t}| \leq c] \\ &\quad + \mathbb{E}\left[Z \Big| |r_{j,t}| > c\right] \mathbb{P}[|r_{j,t}| > c] - a_\psi(c) \\ &= \frac{\nu_j^2}{\kappa^2} \left\{ \mathbb{E}\left[Z\frac{\kappa^2}{\nu_j^2} \Big| |\alpha| \leq \gamma \right] \mathbb{P}[|\alpha| \leq \gamma ] \right\} \\ &\quad + \frac{\nu_j^2}{\kappa^2} \left\{ \mathbb{E}\left[Z\frac{\kappa^2}{\nu_j^2} \Big| |\alpha| > \gamma \right] \mathbb{P}[|\alpha| > \gamma ] \right\} - a_\psi(c). \end{aligned}$$

Denoting $\Phi(\cdot)$ and $\phi(\cdot)$ as being the Gaussian distribution and density functions respectively, using the results of Dhrymes (2005) we have

$$\mathbb{E}[Z - a_\psi(c)] = \frac{c^2}{\gamma^2}(2\Phi(\gamma) - 2\gamma\phi(\gamma) - 1) + 2c^2(1 - \Phi(\gamma)) - a_\psi(c) =$$

$$c^2 \underbrace{\left(\frac{2\Phi(\gamma)}{\gamma^2} - \frac{2\phi(\gamma)}{\gamma} - \frac{1}{\gamma^2} - 2\Phi(\gamma)\right)}_{f(\gamma)} + 2c^2 - a_\psi(c).$$

We define $g(\gamma) := \mathbb{E}[Z - a_\psi(c)]$ which has a unique solution for $\gamma$ if $f(\gamma)$ has a unique solution for $\gamma$. Hence, we focus on $f(\gamma)$ and take its derivative to understand if it is a

strictly monotone function

$$\frac{\partial}{\partial \gamma} f(\gamma) = \frac{2(1 - 2\Phi(\gamma))}{\gamma^3} + \frac{4\phi(\gamma)}{\gamma^2} - \frac{2\phi'(\gamma)}{\gamma} - 2\phi(\gamma)$$

where $\phi'(\gamma) = \frac{\partial}{\partial \gamma} \phi(\gamma) = -\gamma \phi(\gamma)$ which finally gives us

$$\frac{\partial}{\partial \gamma} f(\gamma) = \frac{2}{\gamma^3} \underbrace{(2\gamma\phi(\gamma) + 1 - 2\Phi(\gamma))}_{A}. \tag{B.3}$$

If we prove that the term A in (B.3) is strictly positive or negative, we prove that the derivative is too. By rewriting A we have

$$2\gamma\phi(\gamma) + 1 - 2\Phi(\gamma) = 2\gamma\phi(\gamma) + 2\Phi(0) - 2\Phi(\gamma) = 2(\gamma\phi(\gamma) + \Phi(0) - \Phi(\gamma))$$

and we prove that this is quantity is strictly negative since $\gamma\phi(\gamma) < \Phi(\gamma) - \Phi(0)$ given that $\gamma > 0$.

Now, let us prove the identifiability for the Tukey biweight function. Therefore, in the same manner let

$$Z = \begin{cases} \left(\left(\frac{r_{j,t}}{c}\right)^2 - 1\right)^4 r_{j,t}^2 & \text{for } |r_{j,t}| \leq c \\ 0 & \text{for } |r_{j,t}| > c \end{cases}$$

and let $\kappa$ belong to the set $\{x \in \mathbb{R} \,|\, c^* < x < \infty\}$ where $c^*$ denotes a positive constant such that $c^* < \nu_{0,j}$. Let us again follow the same procedure and notations as used for the proof of identifiability of the Huber weights. With $a_\psi(c)$ being this time the correction term for

the Tukey biweight function, in this case we have

$$\mathbb{E}\left[Z - a_\psi(c)\right] = \mathbb{E}\left[Z\right] - a_\psi(c) = \mathbb{E}\left[Z\,\Big|\,|r_{j,t}| \leq c\right]\mathbb{P}\left[|r_{j,t}| \leq c\right] - a_\psi(c)$$

$$= \frac{\nu_j^{10}}{\kappa^{10} c^8} \left\{\mathbb{E}\left[r_{j,t}^{10} \frac{\kappa^{10}}{\nu_j^{10}}\,\bigg|\,|\alpha| \leq \gamma\right]\mathbb{P}\left[|\alpha| \leq \gamma\right]\right\}$$

$$- \frac{4\nu_j^8}{\kappa^8 c^6} \left\{\mathbb{E}\left[r_{j,t}^8 \frac{\kappa^8}{\nu_j^8}\,\bigg|\,|\alpha| \leq \gamma\right]\mathbb{P}\left[|\alpha| \leq \gamma\right]\right\}$$

$$+ \frac{6\nu_j^6}{\kappa^6 c^4} \left\{\mathbb{E}\left[r_{j,t}^6 \frac{\kappa^6}{\nu_j^6}\,\bigg|\,|\alpha| \leq \gamma\right]\mathbb{P}\left[|\alpha| \leq \gamma\right]\right\}$$

$$- \frac{4\nu_j^4}{\kappa^4 c^2} \left\{\mathbb{E}\left[r_{j,t}^4 \frac{\kappa^4}{\nu_j^4}\,\bigg|\,|\alpha| \leq \gamma\right]\mathbb{P}\left[|\alpha| \leq \gamma\right]\right\}$$

$$+ \frac{\nu_j}{\kappa}\left\{\mathbb{E}\left[r_{j,t}\frac{\kappa}{\nu_j}\,\bigg|\,|\alpha| \leq \gamma\right]\mathbb{P}\left[|\alpha| \leq \gamma\right]\right\} - a_\psi(c)$$

$$= c^2 \Bigg[\frac{1}{\gamma^{10}} \underbrace{(1890\Phi(\gamma) - \overbrace{2\gamma(945 + 315\gamma^2 + 63\gamma^4 + 9\gamma^6 + \gamma^8)\phi(\gamma)}^{\mu_{10}^*} - 945)}_{\mu_{10}}$$

$$- \frac{4}{\gamma^8}\underbrace{(210\Phi(\gamma) - \overbrace{2\gamma(105 + 35\gamma^2 + 7\gamma^4 + \gamma^6)\phi(\gamma)}^{\mu_8^*} - 105)}_{\mu_8}$$

$$+ \frac{6}{\gamma^6}\underbrace{(30\Phi(\gamma) - \overbrace{2\gamma(15 + 5\gamma^2 + \gamma^4)\phi(\gamma)}^{\mu_6^*} - 15)}_{\mu_6}$$

$$+ \frac{4}{\gamma^4}\underbrace{(6\Phi(\gamma) - \overbrace{2\gamma(3 + \gamma^2)\phi(\gamma)}^{\mu_4^*} - 3)}_{\mu_4} + \frac{1}{\gamma^2}\underbrace{(2\Phi(\gamma) - 2\gamma\phi(\gamma) - 1)}_{\mu_2}\Bigg] - a_\psi(c).$$

Next, we define $g(\gamma) := \mathbb{E}\left[Z - a_\psi(c)\right]$ and we know that $g(\gamma)$ has a unique solution in $\gamma$ if the expression in square brackets in $g(\gamma)$ has a unique solution in $\gamma$. Hence, by taking the

derivative we obtain

$$
\begin{aligned}
\frac{\partial}{\partial \gamma} g(\gamma) = & -\frac{\mu_{10}}{\gamma^{11}} + \frac{32}{\gamma^9} \mu_8 + \frac{1}{\gamma^{10}} \big(1890\phi(\gamma) \\
& - \big((1890 + 1890\gamma^2 + 630\gamma^4 + 126\gamma^6 + 18\gamma^8) \phi(\gamma) - \gamma\phi(\gamma)\mu_{10}^*\big)\big) \\
& - \frac{4}{\gamma^8} \big(210\phi(\gamma) - \big((210 + 210\gamma^2 + 70\gamma^4 + 14\gamma^6) \phi(\gamma) - \gamma\phi(\gamma)\mu_8^*\big)\big) \\
& - \frac{36}{\gamma^7} \mu_6 + \frac{6}{\gamma^6} \big(30\phi(\gamma) - \big((30 + 30\gamma^2 + 10\gamma^4) \phi(\gamma) - \gamma\phi(\gamma)\mu_6^*\big)\big) \\
& + \frac{16}{\gamma^5} \mu_4 - \frac{4}{\gamma^4} \big(6\phi(\gamma) - \big((6 + 6\gamma^2) \phi(\gamma) - \gamma\phi(\gamma)\mu_4^*\big)\big) \\
& - \frac{2}{\gamma^3} \mu_2 + \frac{1}{\gamma^2} \big(2\phi(\gamma) - \big(2\phi(\gamma) - 2\gamma^2\phi(\gamma)\big)\big)
\end{aligned}
$$

whose value is strictly negative for $\gamma > 3.5$. The latter condition is very mild since it implies that the equation $\mathbb{E}[\psi(W_{j,t}, \kappa^2)] = 0$ has the unique solution $\kappa^2 = \nu_{0,j}^2$ if $\kappa$ belongs to the set $\{x \in \mathbb{R} \mid 7\nu_{0,j}/2c < x < \infty\}$ for the Tukey biweight function. In other words, the parameter $\nu_{0,j}^2$ is identifiable if $c > 3.5$ so that it belongs to the previously defined set. This condition is very reasonable as it is satisfied for any efficiency larger than approximately 2.5%, an efficiency which is already too low to make any sense in practice.

## C  Proof of Theorem 2.1

In this appendix we discuss the asymptotic normality of the proposed WV estimator $\hat{\boldsymbol{\nu}}$. Before proving Theorem 2.1, we need two additional results which are namely, (i) the consistency of $\hat{\nu}_j^2$ as well as (ii) the Bouligand differentiability of the Huber $\psi$-function if chosen for the estimator $\hat{\boldsymbol{\nu}}$ (which is needed for a MacLaurin expansion in the proof). We start with the consistency of $\hat{\nu}_j^2$ which is stated in the following proposition (followed by its proof).

**Proposition C.1.** *Under Conditions (C2) to (C4), we have that*

$$\hat{\nu}_j^2 \xrightarrow{p} \nu_j^2.$$

9

*Proof.* We firstly verify the point-wise convergence

$$\frac{1}{M_j} \sum_{t=1}^{M_j} \psi(W_{j,t}, \nu_j^2) \xrightarrow{p} \mathbb{E}[\psi(W_{j,t}, \nu_j^2)]$$

for any $\nu_j^2 > 0$. Since $\left(\psi(W_{j,t}, \nu_j^2)\right)$ is a time-invariant function of $(W_{j,t})$, it is also a stationary process (see Wooldridge, 1994) based on Condition (C3). Recalling the notation $\|Z\|_q := (\mathbb{E}[|Z|^q])^{1/q}$ and denoting $A := {}^1\!/\!_{M_j} \sum_{t=1}^{M_j} \psi(W_{j,t}, \nu_j^2) - \mathbb{E}[\psi(W_{j,t}, \nu_j^2)]$, by Markov inequality we have

$$\mathbb{P}\left(|A| \geq \epsilon\right) \leq \frac{\|A\|_2^2}{\epsilon^2}. \tag{C.1}$$

Applying the definition of the projection operator $\mathcal{P}_t \cdot$, computations show that

$$\psi(W_{j,t}, \nu_j^2) - \mathbb{E}[\psi(W_{j,t}, \nu_j^2)] = \sum_{l=0}^{\infty} \mathcal{P}_{t-l} \psi(W_{j,t}, \nu_j^2),$$

hence the numerator on the right side of the inequality in (C.1) can be written as

$$\left\| \frac{1}{M_j} \sum_{t=1}^{M_j} \psi(W_{j,t}, \nu_j^2) - \mathbb{E}[\psi(W_{j,t}, \nu_j^2)] \right\|_2 = \frac{1}{M_j} \left\| \sum_{t=1}^{M_j} \sum_{l=0}^{\infty} \mathcal{P}_{t-l} \psi(W_{j,t}, \nu_j^2) \right\|_2.$$

Noticing that $(\mathcal{P}_{t-l} \psi(W_{j,t}, \nu_j^2))$ (for $t = 1, \ldots, M_j$) forms a martingale difference sequence, by first applying the triangle inequality and then Burkholder's moment inequality for martingale differences (Burkholder, 1988) we have

$$\frac{1}{M_j} \left\| \sum_{t=1}^{M_j} \sum_{l=0}^{\infty} \mathcal{P}_{t-l} \psi(W_{j,t}, \nu_j^2) \right\|_2 \leq \frac{1}{M_j} \sum_{l=0}^{\infty} \left\| \sum_{t=1}^{M_j} \mathcal{P}_{t-l} \psi(W_{j,t}, \nu_j^2) \right\|_2$$

$$\leq \frac{1}{\sqrt{M_j}} \sum_{l=0}^{\infty} \left\| \mathcal{P}_0 \psi(W_{j,l}, \nu_j^2) \right\|_2.$$

We would therefore want to show that the latter term tends to zero to prove consistency.

For this reason, following the proof in Wu (2011) we now write

$$\mathcal{P}_0 \psi(W_{j,t}, \nu_j^2) = \mathbb{E}[\psi(W_{j,t}, \nu_j^2)|\mathcal{F}_0] - \mathbb{E}[\psi(W_{j,t}, \nu_j^2)|\mathcal{F}_{-1}],$$

where, recalling that $W_{j,t}^\star$ is a coupled version of $W_{j,t}$, we can notice that

$$\mathbb{E}[\psi(W_{j,t}, \nu_j^2)|\mathcal{F}_{-1}] = \mathbb{E}[\psi(W_{j,t}^\star, \nu_j^2)|\mathcal{F}_{-1}],$$

since the filtrations $\mathcal{F}_t$ and $\mathcal{F}_t^\star$ are the same up to $t = -1$ (and are different at $t = 0$). This implies that

$$\mathbb{E}[\psi(W_{j,t}, \nu_j^2)|\mathcal{F}_{-1}] = \mathbb{E}[\psi(W_{j,t}^\star, \nu_j^2)|\mathcal{F}_{-1}] = \mathbb{E}[\psi(W_{j,t}^\star, \nu_j^2)|\mathcal{F}_0].$$

This allows us to rewrite

$$\left\|\mathcal{P}_0 \psi(W_{j,t}, \nu_j^2)\right\|_2 = \left\|\mathbb{E}[\psi(W_{j,t}, \nu_j^2) - \psi(W_{j,t}^\star, \nu_j^2)|\mathcal{F}_0]\right\|_2,$$

which, by Jensen's inequality, gives us

$$\left\|\mathbb{E}[\psi(W_{j,t}, \nu_j^2) - \psi(W_{j,t}^\star, \nu_j^2)|\mathcal{F}_0]\right\|_2 \leq \left\|\psi(W_{j,t}, \nu_j^2) - \psi(W_{j,t}^\star, \nu_j^2)\right\|_2.$$

Moreover, recall that

$$\psi(W_{j,t}, \nu_j^2) = \omega^2\left(r_{j,t}; \nu_j^2, c\right) r_{j,t}^2 - a(\nu_j^2, c),$$

where $\omega(\cdot) \in [0, 1]$ are weights (given for example by the Huber or Tukey biweight functions) and $r_{j,t} = W_{j,t}/\nu_j$. Given this, we can denote $\phi(W_{j,t}/\nu_j, c) := \omega\left(r_{j,t}; \nu_j^2, c\right) r_{j,t}$ and,

11

combining the above notations and expansions we have

$$\frac{1}{\sqrt{M_j}} \sum_{l=0}^{\infty} \left\| \mathcal{P}_0 \psi(W_{j,l}, \nu_j^2) \right\|_2$$

$$\leq \frac{1}{\sqrt{M_j}} \sum_{l=0}^{\infty} \left\| \psi(W_{j,l}, \nu_j^2) - \psi(W_{j,l}^\star, \nu_j^2) \right\|_2$$

$$= \frac{1}{\sqrt{M_j}} \sum_{l=0}^{\infty} \left\| [\phi(W_{j,l}/\nu_j, c)]^2 - [\phi(W_{j,l}^\star/\nu_j, c)]^2 \right\|_2$$

$$= \frac{1}{\sqrt{M_j}} \sum_{l=0}^{\infty} \left\| [\phi(W_{j,l}/\nu_j, c) + \phi(W_{j,l}^\star/\nu_j, c)][\phi(W_{j,l}/\nu_j, c) - \phi(W_{j,l}^\star/\nu_j, c)] \right\|_2$$

$$= \frac{1}{\sqrt{M_j}} \sum_{l=0}^{\infty} \mathbb{E} \left\{ [\phi(W_{j,l}/\nu_j, c) + \phi(W_{j,l}^\star/\nu_j, c)]^2 [\phi(W_{j,l}/\nu_j, c) - \phi(W_{j,l}^\star/\nu_j, c)]^2 \right\}^{1/2}.$$

By Hölder's inequality we have that the last term is smaller or equal to

$$\frac{1}{\sqrt{M_j}} \sum_{l=0}^{\infty} 2 \left\| \phi(W_{j,l}/\nu_j, c) \right\|_4 \left\| \phi(W_{j,l}/\nu_j, c) - \phi(W_{j,l}^\star/\nu_j, c) \right\|_4,$$

and, noticing that

$$|\phi(W_{j,t}/\nu_j, c) - \phi(W_{j,t}^\star/\nu_j, c)| \leq \left| \frac{W_{j,t} - W_{j,t}^\star}{\nu_j} \right|,$$

we can finally write

$$\frac{1}{\sqrt{M_j}} \sum_{l=0}^{\infty} \left\| \mathcal{P}_0 \psi(W_{j,l}, \nu_j^2) \right\|_2 \leq \frac{1}{\sqrt{M_j} \nu_j} \sum_{l=0}^{\infty} 2 \left\| \phi(W_{j,l}/\nu_j, c) \right\|_4 \left\| W_{j,l} - W_{j,l}^\star \right\|_4.$$

At this point, we can also underline that $|\phi(W_{j,t}/\nu_j, c)| \leq c$ implying that, for all $p > 0$,

$\|\phi(W_{j,t}/\nu_j, c)\|_p \leq k < \infty$. Using Condition (C4) we finally have

$$\frac{1}{\sqrt{M_j}} \sum_{l=0}^{\infty} \|\mathcal{P}_0 \psi(W_{j,l}, \nu_j^2)\|_2 \leq \frac{2k}{\sqrt{M_j}\nu_j} \sum_{l=0}^{\infty} \delta_{l,4}^j$$
$$= \mathcal{O}_p\left(\frac{1}{\sqrt{M_j}}\right).$$

Therefore, using these results in (C.1) we have

$$\frac{1}{M_j} \sum_{t=1}^{M_j} \psi(W_{j,t}, \nu_j^2) \xrightarrow{p} \mathbb{E}[\psi(W_{j,t}, \nu_j^2)], \tag{C.2}$$

for every $\nu_j^2$. Based on the requirements of Lemma 5.10 of Van der Vaart (2000), knowing that $\mathbf{N} \subset \mathbb{R}^+$ and using Condition (C2) we have that

$$\hat{\nu}_j^2 \xrightarrow{p} \nu_j^2,$$

thus concluding the proof.

□

Having proved consistency, we now deliver an additional technical result that allows to perform an expansion in the case where the Huber $\psi$-function is chosen. This result is provided in the following lemma (followed again by the relative proof).

**Lemma C.1.** *Assuming the wavelet coefficient process $(W_{j,t})$ is Gaussian, the function $\psi(W_{j,t}, \nu_j^2)$ using Huber weights is Bouligand-differentiable as follows*

$$\psi'(W_{j,t}, \nu_j^2) = \begin{cases} -\frac{W_{j,t}^2}{\nu_j^4} & \text{if } |r_{j,t}| \leq c \\ 0 & \text{if } |r_{j,t}| > c \end{cases}$$

The proof of this lemma is given below.

*Proof.* Let us define $r_0 := W_{j,t}/\sqrt{\nu_0^2}$ and $r := W_{j,t}/\sqrt{\nu^2}$ where $\nu^2 = \nu_0^2 + h$. By the definition in Scholtes (2012), a function $f(\cdot)$ is Bouligand differentiable (B-differentiable) at point $x_0$ if it is directionally differentiable at this point and there exists a function $f'(\cdot)$ such that $f(x_0 + h) = f(x_0) + f'(x_0)h + o(h)$. Using the approach of Christmann and Van Messem

13

(2008), we first show that the function $\psi(W_{j,t}, \nu_j^2)$ is first degree B-differentiable using Huber weights. Below are the computations of the B-derivatives for the five cases of the Huber weight function:

1. Setting $r_0 = c$ we have:

   - If $h \geq 0$ ($r \leq c$):

   $$\begin{aligned} \psi'(W_{j,t}, \nu_0^2)h + o(h) &= \psi(W_{j,t}, \nu_0^2 + h) - \psi(W_{j,t}, \nu_0^2) \\ &= r^2 - a_\psi(c) - r_0^2 + a_\psi(c) \\ &= \frac{W_{j,t}^2}{\nu_0^2 + h} - \frac{W_{j,t}^2}{\nu_0^2} \\ &= \frac{W_{j,t}^2}{\nu_0^2}\left(\frac{-h}{\nu_0^2 + h}\right) \\ &= -\frac{W_{j,t}^2}{\nu_0^2}\left(\frac{h}{\nu_0^2} - \frac{h^2}{\nu_0^2(\nu_0^2 + h)}\right) \\ &= -\frac{W_{j,t}^2}{\nu_0^4}h + \underbrace{\frac{W_{j,t}^2 h^2}{\nu_0^2(\nu_0^2 + h)}}_{o(h)} := \Delta \end{aligned}$$

   - If $h < 0$ ($r > c$):

   $$\begin{aligned} \psi'(W_{j,t}, \nu_0^2)(h) + o(h) &= \psi(W_{j,t}, \nu_0^2 + h) - \psi(W_{j,t}, \nu_0^2) \\ &= c^2 - a_\psi(c) - r_0^2 + a_\psi(c) = c^2 - c^2 = 0 \end{aligned}$$

2. Setting $r_0 = -c$ we have:

   - If $h < 0$ ($r < -c$):

   $$\begin{aligned} \psi'(W_{j,t}, \nu_0^2)(h) + o(h) &= \psi(W_{j,t}, \nu_0^2 + h) - \psi(W_{j,t}, \nu_0^2) \\ &= c^2 - a_\psi(c) - r_0^2 + a_\psi(c) = 0 \end{aligned}$$

   - If $h \geq 0$ ($r \geq -c$):

   $$\begin{aligned} \psi'(W_{j,t}, \nu_0^2)(h) + o(h) &= \psi(W_{j,t}, \nu_0^2 + h) - \psi(W_{j,t}, \nu_0^2) \\ &= r^2 - a_\psi(c) - r_0^2 + a_\psi(c) \end{aligned}$$

14

$$= \ldots = \Delta$$

3. Setting $r_0 > c$ we have:

$$\begin{aligned}\psi'(W_{j,t}, \nu_0^2)(h) + o(h) &= \psi(W_{j,t}, \nu_0^2 + h) - \psi(W_{j,t}, \nu_0^2) \\ &= c^2 - a_\psi(c) - r_0^2 + a_\psi(c) = 0\end{aligned}$$

4. Setting $r_0 < -c$ we have:

$$\begin{aligned}\psi'(W_{j,t}, \nu_0^2)(h) + o(h) &= \psi(W_{j,t}, \nu_0^2 + h) - \psi(W_{j,t}, \nu_0^2) \\ &= c^2 - a_\psi(c) - r_0^2 + a_\psi(c) = 0\end{aligned}$$

5. Setting $-c < r_0 < c$ we have:

$$\begin{aligned}\psi'(W_{j,t}, \nu_0^2)(h) + o(h) &= \psi(W_{j,t}, \nu_0^2 + h) - \psi(W_{j,t}, \nu_0^2) \\ &= r^2 - a_\psi(c) - r_0^2 + a_\psi(c) \\ &= \ldots = \Delta\end{aligned}$$

We therefore have that the first B-derivative of the function $\psi(W_{j,t}, \nu_j^2)$ is given by

$$\psi'(W_{j,t}, \nu_j^2) = \begin{cases} -\frac{W_{j,t}^2}{\nu_j^4} & \text{if } |r_{j,t}| \leq c \\ 0 & \text{if } |r_{j,t}| > c \end{cases}$$

The approach used in this proof can be used to obtain expressions for the B-derivatives of other piecewise differentiable weight functions (see Scholtes, 2012). It can be seen how it extends the classic derivative for $|r_0| < c$ also to the points $\nu_0^2$ such that $|r_0| = c$. However, the Frechet differentiability of this function has also been discussed in Clarke (1986). □

As mentioned earlier, Lemma C.1 is useful for the results on asymptotic normality of the proposed estimator to hold in case the choice of the $\psi$-function corresponds to the Huber $\psi$-function, proving that this function respects Condition (C1) which is required by Theorem 2.1. The proof of this theorem, which is valid for a general $\psi$-function that respects Condition (C1), is given below.

*Proof.* Given Condition (C1), let us denote $\psi'(W_{j,t}, \nu_j{}^2) = \partial/\partial\nu_j{}^2\, \psi(W_{j,t}, \nu_j{}^2)$ and apply the mean value theorem to $\sum_{t=1}^{M_j} \psi(W_{j,t}, \hat{\nu}_j^2)$ around $\nu_j^2$ obtaining

$$\sum_{t=1}^{M_j} \psi(W_{j,t}, \hat{\nu}_j^2) = \sum_{t=1}^{M_j} \psi(W_{j,t}, \nu_j^2) + \sum_{t=1}^{M_j} \psi'(W_{j,t}, \nu_j^{*2})(\hat{\nu}_j^2 - \nu_j^2) = 0$$

where

$$|\nu_j^{*2} - \nu_j^2| \leq |\hat{\nu}_j^2 - \nu_j^2|. \tag{C.3}$$

Rearranging the expansion and multiplying by $\sqrt{T}$ yields

$$\sqrt{T}(\hat{\nu}_j^2 - \nu_j^2) = \sqrt{\frac{T}{M_j}} \underbrace{\left[-\frac{1}{M_j}\sum_{t=1}^{M_j}\psi'(W_{j,t}, \nu_j^{*2})\right]^{-1}}_{A_j} \underbrace{\frac{1}{\sqrt{M_j}}\sum_{t=1}^{M_j}\psi(W_{j,t}, \nu_j^2)}_{B_j}. \tag{C.4}$$

Let us start from term $A_j$. We can rewrite this term as

$$-\frac{1}{M_j}\sum_{t=1}^{M_j}\psi'(W_{j,t}, \nu_j^{*2}) = -\frac{1}{M_j}\sum_{t=1}^{M_j}\psi'(W_{j,t}, \nu_j^2) - \frac{1}{M_j}\sum_{t=1}^{M_j}\underbrace{[\psi'(W_{j,t}, \nu_j^{*2}) - \psi'(W_{j,t}, \nu_j^2)]}_{C_j}.$$

Since $\bigl(\psi'(W_{j,t}, \nu_j^2)\bigr)$ is a time-invariant function of $(W_{j,t})$, it is also a stationary process (see Wooldridge, 1994) based on Condition (C3). Let us start from the first term on the right side of the above equality and define $m_j = \mathbb{E}[-\psi'(W_{j,t}, \nu_j^2)]$. Then by Markov inequality we have

$$\mathbb{P}\left(\left|\frac{1}{M_j}\sum_{t=1}^{M_j}-\psi'(W_{j,t}, \nu_j^2) - m_j\right| \geq \epsilon\right) \leq \frac{\left\|\frac{1}{M_j}\sum_{t=1}^{M_j}-\psi'(W_{j,t}, \nu_j^2) - m_j\right\|_2^2}{\epsilon^2},$$

where, following the same reasoning as for the proof of Proposition C.1, $\left\| \frac{1}{M_j} \sum_{t=1}^{M_j} -\psi'(W_{j,t}, \nu_j^2) - m_j \right\|_2^2$ can be bounded by following inequalities,

$$\left\| \frac{1}{M_j} \sum_{t=1}^{M_j} -\psi'(W_{j,t}, \nu_j^2) - m_j \right\|_2^2 = \frac{1}{M_j^2} \left\| \sum_{t=1}^{M_j} \sum_{l=0}^{\infty} \mathcal{P}_{t-l} \psi'(W_{j,t}, \nu_j^2) \right\|_2^2$$

$$\leq \frac{1}{M_j^2} \sum_{l=0}^{\infty} \left\| \sum_{t=1}^{M_j} \mathcal{P}_{t-l} \psi'(W_{j,t}, \nu_j^2) \right\|_2^2$$

$$\leq \frac{1}{M_j} \sum_{l=0}^{\infty} \left\| \mathcal{P}_0 \psi'(W_{j,l}, \nu_j^2) \right\|_2^2.$$

Under Conditions (C1) to (C3), we can follow the same steps as the proof of Proposition C.1 to deliver

$$\frac{1}{M_j} \sum_{t=1}^{M_j} -\psi'(W_{j,t}, \nu_j^2) \xrightarrow{p} \mathbb{E}[-\psi'(W_{j,t}, \nu_j^2)] = m_j.$$

As for term $C_j$, since $\hat{\nu}_j^2$ is a consistent estimator of $\nu_j^2$, by (C.3) so is $\nu_j^{*2}$. Moreover, since $\psi'(\cdot)$ is continuous almost everywhere by Condition (C1), using the continuous mapping theorem we have

$$\frac{1}{M_j} \sum_{t=1}^{M_j} [\psi'(W_{j,t}, \nu_j^{*2}) - \psi'(W_{j,t}, \nu_j^2)] \xrightarrow{p} 0,$$

which finally yields

$$A_j \xrightarrow{p} m_j.$$

Let us now focus on term $B_j$ for which we intend to show convergence to a normal distribution in order to make use of Slutsky's theorem. For this reason we verify the requirements of Theorem 7 in Wu (2011) most of which have already been verified in the proof of Proposition C.1. Indeed, based Condition (C3) we have that $(\psi(W_{j,t}, \nu_j^2))$ is a stationary process

which can be represented as

$$\psi(W_{j,t}, \nu_j^2) = \sum_{l=0}^{\infty} \mathcal{P}_{t-l}\psi(W_{j,t}, \nu_j^2),$$

where $\left(\mathcal{P}_{t-l}\psi(W_{j,t}, \nu_j^2)\right)_{l=0,\ldots,\infty}$ is a martingale difference sequence. Based on this, we verify the conditions of Theorem 3 in Wu (2011) such that the martingale central limit theorem can be applied. Firstly we need to show that

$$\sum_{l=0}^{\infty} \left\| \mathcal{P}_0\psi(W_{j,l}, \nu_j^2) \right\|_2^2 < \infty.$$

This requirement was verified in the proof of Proposition C.1 since it was shown that, with $k < \infty$, we have

$$\sum_{l=0}^{\infty} \left\| \mathcal{P}_0\psi(W_{j,l}, \nu_j^2) \right\|_2 \leq \frac{2\,k}{\nu_j} \sum_{l=0}^{\infty} \delta_{l,4}^j,$$

thereby, based on Condition (C4), verifying the above requirement. Hence, based on Theorem 3 in Wu (2011) we have that the term $B_j$ has the following asymptotic distribution

$$\frac{1}{\sqrt{M_j}} \sum_{t=1}^{M_j} \psi(W_{j,t}, \nu_j^2) \xrightarrow{\mathcal{D}} \mathcal{N}\left(0, \mathbb{E}[D_0^2]\right),$$

where $D_0 := \sum_{t=0}^{\infty} \mathcal{P}_0\psi(W_{j,t}, \nu_j^2)$. Since $M_j = \mathcal{O}(T)$, using all the above results we apply Slutsky's theorem to (C.4) to obtain

$$\sqrt{T}(\hat{\nu}_j^2 - \nu_j^2) \xrightarrow{\mathcal{D}} \mathcal{N}\left(0, \frac{\mathbb{E}[D_0^2]}{m_j^2}\right).$$

Finally, employing the Crámer-Wold device we can deliver the final result:

$$\sqrt{T}\left(\hat{\boldsymbol{\nu}} - \boldsymbol{\nu}\right) \xrightarrow{\mathcal{D}} \mathcal{N}\left(\mathbf{0}, \boldsymbol{V}\right),$$

where $\boldsymbol{V} = \boldsymbol{M}\,\mathbb{E}[\boldsymbol{D}_0\boldsymbol{D}_0^\intercal]\,\boldsymbol{M}^\intercal$ with $\boldsymbol{D}_0 := \sum_{t=0}^{\infty} \mathcal{P}_0\psi(\boldsymbol{W}_t, \boldsymbol{\nu})$ and $\boldsymbol{M} := \mathbb{E}[-\partial/\partial\boldsymbol{\nu}\,\psi(\boldsymbol{W}_t, \boldsymbol{\nu})]$
. □

# D Further Investigation of Model Identifiability using the Wavelet Variance

This section further studies the identifiability results presented in Guerrier et al. (2013) and, for this purpose, makes use of the conditions required in Komunjer (2012). These conditions (or assumptions) are the following:

(A1) The function $\boldsymbol{\nu}(\boldsymbol{\theta})$ is twice continuously differentiable.

(A2) $|\boldsymbol{\nu}(\boldsymbol{\theta})|_2 \to \infty$ whenever $|\boldsymbol{\theta}|_2 \to \infty$.

(A3) For every $\boldsymbol{\theta} \in \mathbb{R}^p$, the Jacobian of $\boldsymbol{\nu}(\boldsymbol{\theta})$ is nonnegative (or non-positive).

(A4) For every $\boldsymbol{c} \in \mathbb{R}^p$ the equation $\boldsymbol{\nu}(\boldsymbol{\theta}) = \boldsymbol{c}$ has countably (possibly zero) solutions in $\mathbb{R}^p$.

In order to study model identifiability we denote a process as $\left(X_t^{(j)}\right)$ with $t = 1\ldots,T$ ($T \in \mathbb{N}_+$) and $j$ indicating a specific model that characterizes this process. With this notation, we define the following time series models:

(T1) <u>White Noise</u> (WN) with parameter $\sigma^2 \in \mathbb{R}^+$. We denote this process as $\left(X_t^{(1)}\right)$.

(T2) <u>Quantization Noise</u> (QN) (or rounding error, see e.g. Papoulis, 1991) with parameter $Q^2 \in \mathbb{R}^+$. We denote this process as $\left(X_t^{(2)}\right)$.

(T3) <u>Drift</u> with parameter $\omega \in \mathbb{R}^+$ (or $\omega \in \mathbb{R}^-$). We denote this process as $\left(X_t^{(3)}\right)$.

(T4) <u>Random walk</u> (RW) with parameter $\gamma^2 \in \mathbb{R}^+$. We denote this process as $\left(X_t^{(4)}\right)$.

(T5) <u>Moving Average</u> (MA(1)) process with non-zero parameter $\varrho \in (-1, +1)$ and $\varsigma^2 \in \mathbb{R}^+$. We denote this process as $\left(X_t^{(5)}\right)$.

(T6) <u>Auto-Regressive</u> (AR(1)) process with parameters $\rho_k \in (-1, +1)$ and $\upsilon_k^2 \in \mathbb{R}^+$ such that $\rho_k < \rho_{k'}, \forall k < k'$. We denote this process as $\left(X_t^{(j)}\right), j = 6,\ldots,G$ with $G \in \mathbb{N}^+$, $6 \leq G < \infty$.

Processes (T1), (T2), (T5) and (T6) are stationary models based on their parameter definitions. Process (T2) is particularly useful in the field of engineering where measurements are often rounded therefore introducing an error process in the procedure. In general terms, this process can be represented as a linear combination of differences of standard uniform variables where $Q^2$ plays the role of scaling factor. Processes (T3) and (T4) are typically non-stationary processes, where process (T3) is a non-random linear function with slope $\omega$. Focusing on the commonly used Haar wavelet filter, the theoretical WV $\boldsymbol{\nu}(\boldsymbol{\theta})$ for these models can be found, for example, in Zhang (2008) where these expressions are given for the Allan variance (i.e. twice the Haar WV). Using this notation, we will firstly focus on some classes of (latent) time series models given below:

**Model 1** $W_t = \sum_{i=6}^{G} X_t^{(i)}$.

**Model 2** $W_t = \sum_{i=1}^{4} X_t^{(i)}$.

**Model 3** $W_t = \sum_{i=3}^{5} X_t^{(i)}$.

**Model 1** is a stationary latent model since it consists of a sum of $K = G - 5$ (T6) processes which, based on the results in Granger and Morris (1976), can be interpreted as a reparametrization of ARMA models. On the other hand, **Model 2** and **Model 3** are classes of latent models that combine stationary and non-stationary processes.

For all the above models we know that Assumption **(A1)** is respected while, in order to respect Assumption **(A2)**, we consider an injective transformation of their parameters. Indeed, for all parameters that belong to the positive part of the real line (e.g. variance parameters) we can choose a log-transformation while for all parameters that belong to $(-1, 1)$ (e.g. autorgressive and moving average parameters) we can use the tangent function. For example, assuming that $\tilde{\boldsymbol{\theta}} = [\tilde{\rho} \; \tilde{\upsilon}^2] \in \mathbb{R}^2$, the parameters of a first-order autoregressive process (T6) can be obtained as follows:

$$\rho = \frac{2}{\pi} \arctan \tilde{\rho}$$

and

$$\upsilon^2 = \exp(\tilde{\upsilon}^2).$$

Hence, for the models considered above we can always find an injective transformation such that Assumption **(A2)** is respected. While this transformation preserves the validity of Assumption **(A1)**, we need to verify that Assumptions **(A3)** and **(A4)** hold as well when considering the parameter vector $\tilde{\boldsymbol{\theta}} \in \mathbb{R}^p$.

20

In order to study Assumptions **(A3)** and **(A4)** we will compute the determinant of the Jacobian of $\boldsymbol{\nu}(\tilde{\boldsymbol{\theta}})$ and, to do so, we will make use of the following chain rule where we denote $\boldsymbol{f}(\cdot)$ as a generic (injective) function (such as those mentioned above) applied to the vector $\tilde{\boldsymbol{\theta}}$ and $\boldsymbol{J}$ as being the Jacobian:

$$\boldsymbol{J}_{\boldsymbol{\nu}(\boldsymbol{f}(\cdot))}(\tilde{\boldsymbol{\theta}}) = \underbrace{\boldsymbol{J}_{\boldsymbol{\nu}(\cdot)}\left(\boldsymbol{f}(\tilde{\boldsymbol{\theta}})\right)}_{A} \underbrace{\boldsymbol{J}_{\boldsymbol{f}(\cdot)}(\tilde{\boldsymbol{\theta}})}_{D}.$$

Based on this, we know that the determinant of $\boldsymbol{J}_{\boldsymbol{\nu}(\boldsymbol{f}(\cdot))}(\tilde{\boldsymbol{\theta}})$ is given by the product of the determinants of matrices A and D. If we prove that the determinant of $\boldsymbol{J}_{\boldsymbol{\nu}(\boldsymbol{f}(\cdot))}(\tilde{\boldsymbol{\theta}})$ is either strictly positive or strictly negative, then we simultaneously prove Assumptions **(A3)** and **(A4)**.

Let us start by studying the determinant of the Jacobian matrix D. In all cases, each transformation only depends on one distinct element of $\tilde{\boldsymbol{\theta}}$ implying that the resulting Jacobian is diagonal where, using the autoregressive parameters as an example, the elements on this diagonal are of the following form

$$\beta_1 := \frac{\partial}{\partial \tilde{\rho}} \boldsymbol{f}(\tilde{\boldsymbol{\theta}}) = \frac{2}{\pi} \frac{1}{1+\tilde{\rho}^2}$$

$$\beta_2 := \frac{\partial}{\partial \tilde{v}^2} \boldsymbol{f}(\tilde{\boldsymbol{\theta}}) = \exp(\tilde{v}^2).$$

Noticing that both derivative expressions are strictly positive, the form of matrix D is as follows:

$$D = \begin{pmatrix} \beta_j^1 & 0 & \cdots & \cdots & 0 \\ 0 & \beta_j^2 & \cdots & \cdots & \vdots \\ \vdots & \vdots & \ddots & \vdots & \vdots \\ \vdots & \cdots & \cdots & \beta_j^{p-1} & 0 \\ 0 & \cdots & \cdots & 0 & \beta_j^p \end{pmatrix},$$

where $\beta_j^i$ (for $j = 1, 2$) represents the derivative with respect to the $i^{th}$ element of $\tilde{\boldsymbol{\theta}}$. In this case the determinant is given by the product of the diagonal elements (since it is a diagonal matrix) which are all strictly positive implying that its determinant is strictly positive.

Given this result for the Jacobian D, let us analyse the determinant for Jacobian A which, as opposed to D, is strictly model-dependent. For this reason, let us start from the Jacobian A for **Model 1** (i.e. the sum of $K$ first-order autogressive processes) which can

be reparametrized as an ARMA model. For $K \leq 4$, taking the first $2 \times K$ scales of the vector $\boldsymbol{\nu}(\tilde{\boldsymbol{\theta}})$ whose form can be found in Zhang (2008) we are able to directly verify that the determinant of A for this model is given by

$$\det(A) = \frac{\prod_{i=1}^{K} v_i^2 \prod_{i<j}^{K}(\rho_i - \rho_j)^4}{\prod_{i=1}^{K}(\rho_i^2 - 1)^2},$$

which, following the parameter definition for (T6), is always strictly positive. Based on this result we can conjecture that the determinant is of the above form for all sums of (T6) processes with $K < \infty$. As a consequence, all models up to an ARMA(4,3) that can be reparametrized as a sum of (T6) processes fulfill Assumptions **(A3)** and **(A4)** since the determinant of the matrix $\boldsymbol{J}_{\boldsymbol{\nu}(\boldsymbol{f}(\cdot))}(\tilde{\boldsymbol{\theta}})$ is strictly positive based on the product of A and D. These results therefore verify Assumptions **(A1)** to **(A4)** and prove that the sum of $K \leq 4$ (T6) processes is identifiable and consequently so are ARMA models (up to order (4,3)) that can be uniquely reparametrized as the sum of $K$ first-order autoregressive models (and we conjecture that this is the case also for $K > 4$ and hence also for higher order ARMA models that can be reparametrized as a sum of first-order autoregressive models).

**Remark D.1** (ARMA model reparametrization). *It is not certain that the ARMA models that can be reparametrized as a sum of first-order autoregressive models can also be uniquely reparametrized (i.e. there exists a unique mapping between the parameters of the sum of (T6) processes and those of the ARMA model). An example of such a reparametrization is given in Hamilton (1994) (Equation 4.7.26, Section 4.7) where, denoting the parameters of an ARMA(2,1) process as $\bar{\boldsymbol{\theta}} = [\bar{\rho}_1 \ \bar{\rho}_2 \ \bar{\varrho} \ \bar{v}^2]^\intercal$, the latter parameters can be represented as follows*

$$\bar{\boldsymbol{\theta}} = \begin{bmatrix} \bar{\rho}_1 \\ \bar{\rho}_2 \\ \bar{\varrho} \\ \bar{v}^2 \end{bmatrix} = \begin{bmatrix} \rho_1 + \rho_2 \\ -\rho_1 \rho_2 \\ \frac{\rho_1 v_1^2 + \rho_2 v_2^2}{v_1^2 + v_2^2} \\ v_1^2 + v_2^2 \end{bmatrix} = \boldsymbol{g}(\boldsymbol{\theta})$$

*where $\boldsymbol{\theta} = [\rho_1 \ v_1^2 \ \rho_2 \ v_2^2]^\intercal$ is the vector of parameters for the sum of two (T6) processes. Using the same steps as above, we can transform the parameters $\boldsymbol{\theta}$ to obtain $\tilde{\boldsymbol{\theta}}$ and verify Assumptions **(A1)** and **(A2)**. By taking the Jacobian $A = \partial/\partial \tilde{\boldsymbol{\theta}} \ \boldsymbol{g}(\tilde{\boldsymbol{\theta}})$ and applying the reasoning used in the above discussion, we have that the determinant of this matrix is given*

*by*

$$det(A) = \frac{(\tilde{\rho}_1 - \tilde{\rho}_2)^2}{\tilde{v}_1^2 + \tilde{v}_2^2}$$

*which is always positive thereby implying that the matrix A is of full rank. Since the Jacobian of the transformation of the parameters is also strictly positive we verify Assumptions (A3) and (A4) and conclude that there is a unique mapping from the parameters of a sum of two (T6) processes to those of an ARMA(2,1) process. This does not prove the unique mapping of all ARMA models with the sum of (T6) models but supports the assumption that this could be the case for other models.*

Having studied the identifiability for **Model 1** (e.g. ARMA models), let us now discuss the identifiability for **Model 2** and **Model 3** which include non-stationary time series models. Given that Assumptions **(A1)** and **(A2)** are easily verified also for these models, let us verify the determinant of matrix A using the expression for $\boldsymbol{\nu}(\tilde{\boldsymbol{\theta}})$ given in Zhang (2008) (since D has strictly positive determinant by definition). For **Model 2** we have that the determinant of A (taking the first four scales of the vector $\boldsymbol{\nu}(\tilde{\boldsymbol{\theta}})$) is given by $\det(A) = {}^{2205\omega}/_{4096}$ which is either strictly negative or strictly positive depending on the value of $\omega$. Using the same approach for **Model 3** we have that the determinant is given by $\det(A) = -{}^{2205\omega\varrho}/_{256}$ which is also either strictly negative or strictly positive given the parameter definitions. Hence, all assumptions are verified thereby proving identifiability for **Model 2** and **Model 3**.

**Remark D.2** (General Identifiability). *In general, identifiability is a model-specific topic and is therefore proven on a case-by-case basis. However, the above discussion delivered a further study of model identifiability through the (R)GMWM in addition to the results in Guerrier et al. (2013) for certain (general) classes of models which include non-stationary models. Moreover, the study of Greenhall (1998) on the Allan variance (which is equal to the Haar WV up to a constant) highlights how there are only extreme cases where the Allan variance does not benefit from a unique mapping with the spectral density in the continuous-time case. Although these results cannot be directly applied to the discrete-time case of this paper, they constitute an argument supporting the identifiability of the (Haar) WV through the spectral density function. In this case, any model that can be identified through the autocovariance function (and hence the spectral density function) could also be identified through the WV.*

# E  RGMWM Asymptotic Properties

The results of the proofs in this appendix largely follow the conditions and steps for standard extremum estimators (see Newey and McFadden, 1994).

## E.1  RGMWM Consistency

*Proof.* Recall that the $L_q$-norm is denoted as $|\boldsymbol{a}|_q := \left(\sum_{j=1}^{p} |a_j|^q\right)^{1/q}$, where $\boldsymbol{a} = (a_1, \ldots, a_p)^\intercal \in \mathbb{R}^p$, and that $\|\boldsymbol{X}\|_S$ denotes the spectral norm of a matrix $\boldsymbol{X}$. Now let

$$Q(\boldsymbol{\theta}) := (\boldsymbol{\nu}(\boldsymbol{\theta}_0) - \boldsymbol{\nu}(\boldsymbol{\theta}))^\intercal \boldsymbol{\Omega} (\boldsymbol{\nu}(\boldsymbol{\theta}_0) - \boldsymbol{\nu}(\boldsymbol{\theta})) = |\boldsymbol{\nu}(\boldsymbol{\theta}_0) - \boldsymbol{\nu}(\boldsymbol{\theta})|_{\boldsymbol{\Omega}}^2$$

and

$$Q_T(\boldsymbol{\theta}) := (\hat{\boldsymbol{\nu}} - \boldsymbol{\nu}(\boldsymbol{\theta}))^\intercal \hat{\boldsymbol{\Omega}} (\hat{\boldsymbol{\nu}} - \boldsymbol{\nu}(\boldsymbol{\theta})) = |\hat{\boldsymbol{\nu}} - \boldsymbol{\nu}(\boldsymbol{\theta})|_{\hat{\boldsymbol{\Omega}}}^2$$

and define $\boldsymbol{\Omega}^* = \hat{\boldsymbol{\Omega}} - \boldsymbol{\Omega}$. By Theorem 2.1. of Newey and McFadden (1994) we want to prove that $Q_T(\boldsymbol{\theta})$ converges uniformly in probability to $Q(\boldsymbol{\theta})$. By the triangular inequality we have that

$$\left|Q_T(\boldsymbol{\theta}) - Q(\boldsymbol{\theta})\right| \leq \left|\underbrace{|\boldsymbol{\nu}(\boldsymbol{\theta}_0) - \boldsymbol{\nu}(\boldsymbol{\theta})|_{\boldsymbol{\Omega}^*}^2}_{a_1}\right| + \left|\underbrace{|\hat{\boldsymbol{\nu}} - \boldsymbol{\nu}(\boldsymbol{\theta}_0)|_{\hat{\boldsymbol{\Omega}}}^2}_{a_2}\right|$$
$$+ \left|\underbrace{2\left(\boldsymbol{\nu}(\boldsymbol{\theta}_0) - \boldsymbol{\nu}(\boldsymbol{\theta})^\intercal \hat{\boldsymbol{\Omega}} (\hat{\boldsymbol{\nu}} - \boldsymbol{\nu}(\boldsymbol{\theta}_0))\right)}_{a_3}\right|.$$

Considering term $a_1$ and using the same inequalities, we have

$$a_1 \leq \sup_{\boldsymbol{\theta} \in \boldsymbol{\Theta}} |\boldsymbol{\nu}(\boldsymbol{\theta}_0) - \boldsymbol{\nu}(\boldsymbol{\theta})|_2^2 \, \|\boldsymbol{\Omega}^*\|_S.$$

By Condition (C5) we have that $|\boldsymbol{\nu}(\boldsymbol{\theta}_0) - \boldsymbol{\nu}(\boldsymbol{\theta})|_2^2$ is bounded, say by $B < \infty$, and using Condition (C8) we have that

$$a_1 \leq JB \, \|\boldsymbol{\Omega}^*\|_S \xrightarrow{p} 0.$$

Based on Condition (C8) we can also state that $\|\hat{\boldsymbol{\Omega}}\|_S \leq \lambda < \infty$. Moreover, using the results in Proposition C.1 we have that $|\hat{\boldsymbol{\nu}} - \boldsymbol{\nu}(\boldsymbol{\theta}_0)|_2 \xrightarrow{p} 0$. Therefore, again using Condition

**(C8)** we have
$$a_2 \leq \lambda |\hat{\boldsymbol{\nu}} - \boldsymbol{\nu}(\boldsymbol{\theta}_0)|_2^2 \xrightarrow{P} 0.$$

Finally, we have
$$a_3 \leq \sup_{\boldsymbol{\theta} \in \Theta} 2 \|\hat{\boldsymbol{\Omega}}\|_S |\boldsymbol{\nu}(\boldsymbol{\theta}_0) - \boldsymbol{\nu}(\boldsymbol{\theta})|_2 |\hat{\boldsymbol{\nu}} - \boldsymbol{\nu}(\boldsymbol{\theta}_0)|_2,$$

which, using the conditions and arguments for the previous terms, goes to zero based on the boundedness of $\|\hat{\boldsymbol{\Omega}}\|_S$ and $|\boldsymbol{\nu}(\boldsymbol{\theta}_0) - \boldsymbol{\nu}(\boldsymbol{\theta})|_2$ and on the consistency of $\hat{\boldsymbol{\nu}}$. Therefore, we have that
$$\sup_{\boldsymbol{\theta} \in \Theta} |Q_T(\boldsymbol{\theta}) - Q(\boldsymbol{\theta})| \xrightarrow{P} 0.$$

Based on Condition **(C7)** and the non-singularity of $\boldsymbol{\Omega}$ we have that $Q(\boldsymbol{\theta})$ has a unique minimum at $\boldsymbol{\theta}_0$. Therefore, using Theorem 2.1. of Newey and McFadden (1994), by Conditions **(C5)**, **(C6)** and **(C8)** we finally have
$$\hat{\boldsymbol{\theta}} \xrightarrow{P} \boldsymbol{\theta}_0$$

thus concluding the proof. □

## E.2 RGMWM Asymptotic Normality

*Proof.* Given the results on the consistency in Proposition 3.1, the proof of asymptotic normality of $\hat{\boldsymbol{\theta}}$ naturally follows the standard proof of asymptotic normality for extremum estimators. Here we therefore simply give a quick overview of the steps which are based on existing results (see Newey and McFadden, 1994).

Under Condition **(C6)**, by the definition of $\hat{\boldsymbol{\theta}}$ we have

$$\left. \frac{\partial Q_T(\boldsymbol{\theta})}{\partial \boldsymbol{\theta}} \right|_{\boldsymbol{\theta} = \hat{\boldsymbol{\theta}}} = \mathbf{0}_{p \times 1}$$
$$\iff \left. \frac{\partial}{\partial \boldsymbol{\theta}} \left[ (\hat{\boldsymbol{\nu}} - \boldsymbol{\nu}(\boldsymbol{\theta}))^\intercal \hat{\boldsymbol{\Omega}} (\hat{\boldsymbol{\nu}} - \boldsymbol{\nu}(\boldsymbol{\theta})) \right] \right|_{\boldsymbol{\theta} = \hat{\boldsymbol{\theta}}} = \mathbf{0}_{p \times 1}$$

which, up to a constant, yields

$$\underbrace{\left( \frac{\partial}{\partial \boldsymbol{\theta}} (\hat{\boldsymbol{\nu}} - \boldsymbol{\nu}(\boldsymbol{\theta}))^\intercal |_{\boldsymbol{\theta} = \hat{\boldsymbol{\theta}}} \right)}_{\boldsymbol{B}(\hat{\boldsymbol{\theta}})} \hat{\boldsymbol{\Omega}} \left( \hat{\boldsymbol{\nu}} - \boldsymbol{\nu}(\hat{\boldsymbol{\theta}}) \right) = \mathbf{0}_{p \times 1}. \tag{E.1}$$

The multivariate mean value theorem ensures that, based on Condition (C9), there exists a matrix $A(\hat{\boldsymbol{\theta}}, \boldsymbol{\theta}_0)$ that can be used to expand $\hat{\boldsymbol{\nu}} - \boldsymbol{\nu}(\hat{\boldsymbol{\theta}})$ around $\boldsymbol{\theta}_0$ in the following way

$$\hat{\boldsymbol{\nu}} - \boldsymbol{\nu}(\hat{\boldsymbol{\theta}}) = \hat{\boldsymbol{\nu}} - \boldsymbol{\nu}(\boldsymbol{\theta}_0) + \boldsymbol{A}(\hat{\boldsymbol{\theta}}, \boldsymbol{\theta}_0)\left(\hat{\boldsymbol{\theta}} - \boldsymbol{\theta}_0\right). \tag{E.2}$$

Since $\hat{\boldsymbol{\nu}} \xrightarrow{p} \boldsymbol{\nu}(\boldsymbol{\theta}_0)$ and $\hat{\boldsymbol{\theta}} \xrightarrow{p} \boldsymbol{\theta}_0$, the multivariate mean value theorem also guarantees that the matrix $\boldsymbol{A}(\hat{\boldsymbol{\theta}}, \boldsymbol{\theta}_0)$ has the following property

$$\boldsymbol{A}(\hat{\boldsymbol{\theta}}, \boldsymbol{\theta}_0) \xrightarrow{p} \frac{\partial}{\partial \boldsymbol{\theta}^\intercal}\left(\boldsymbol{\nu}(\boldsymbol{\theta}_0) - \boldsymbol{\nu}(\boldsymbol{\theta})\right)\bigg|_{\boldsymbol{\theta}=\boldsymbol{\theta}_0} = \frac{\partial}{\partial \boldsymbol{\theta}^\intercal}\boldsymbol{\nu}(\boldsymbol{\theta})\bigg|_{\boldsymbol{\theta}=\boldsymbol{\theta}_0},$$

given that $\partial/\partial\boldsymbol{\theta}\,\boldsymbol{\nu}(\boldsymbol{\theta}^\intercal)$ is continuous. Plugging (E.2) in the third factor of (E.1), multiplying by $\sqrt{T}$ and using Condition (C10) gives us

$$\sqrt{T}\left(\hat{\boldsymbol{\theta}} - \boldsymbol{\theta}_0\right) = -\left[\boldsymbol{B}(\hat{\boldsymbol{\theta}})\,\hat{\boldsymbol{\Omega}}\,\boldsymbol{A}(\hat{\boldsymbol{\theta}}, \boldsymbol{\theta}_0)\right]^{-1} \boldsymbol{B}(\hat{\boldsymbol{\theta}})\,\hat{\boldsymbol{\Omega}}\,\sqrt{T}\left(\hat{\boldsymbol{\nu}} - \boldsymbol{\nu}(\boldsymbol{\theta}_0)\right). \tag{E.3}$$

Knowing that $\boldsymbol{B}(\hat{\boldsymbol{\theta}}) \xrightarrow{p} \boldsymbol{A}(\boldsymbol{\theta}_0)^\intercal$ by the continuous mapping theorem, by Slutsky's theorem we have that

$$\left[\boldsymbol{B}(\hat{\boldsymbol{\theta}})\,\hat{\boldsymbol{\Omega}}\,\boldsymbol{A}(\hat{\boldsymbol{\theta}}, \boldsymbol{\theta}_0)\right]^{-1} \boldsymbol{B}(\hat{\boldsymbol{\theta}})\,\hat{\boldsymbol{\Omega}} \xrightarrow{p} \left[\boldsymbol{A}(\boldsymbol{\theta}_0)^\intercal \boldsymbol{\Omega} \boldsymbol{A}(\boldsymbol{\theta}_0)\right]^{-1} \boldsymbol{A}(\boldsymbol{\theta}_0)^\intercal \boldsymbol{\Omega}$$

By again using Slutsky's theorem as well Theorem 2.1 we have that (E.3) has the following asymptotic distribution

$$\sqrt{T}\left(\hat{\boldsymbol{\theta}} - \boldsymbol{\theta}_0\right) \xrightarrow[T\to\infty]{\mathcal{D}} \mathcal{N}(\boldsymbol{0}, \boldsymbol{\Xi}),$$

where, denoting $\boldsymbol{H}(\boldsymbol{\theta}_0) := [\boldsymbol{A}(\boldsymbol{\theta}_0)^\intercal \boldsymbol{\Omega} \boldsymbol{A}(\boldsymbol{\theta}_0)]$, the asymptotic covariance matrix $\boldsymbol{\Xi}$ is given by

$$\boldsymbol{\Xi} := \boldsymbol{H}(\boldsymbol{\theta}_0)^{-1} \boldsymbol{A}(\boldsymbol{\theta}_0)^\intercal \boldsymbol{\Omega} \boldsymbol{V} \boldsymbol{\Omega} \boldsymbol{A}(\boldsymbol{\theta}_0) \boldsymbol{H}(\boldsymbol{\theta}_0)^{-1}.$$

and $\boldsymbol{V}$ is given in Theorem 2.1

$\square$

# F   Additional Simulation Studies

## F.1   Wavelet Variance Estimation

In this appendix we investigate the performance of the proposed M-estimator of WV in (2.3) which we denote as RWV in this section . For this purpose, we compare it with the standard estimator of WV, denoted as CL, and with the the robust estimator proposed in Mondal and Percival (2012), denoted as MP. With respect to the latter estimator, we implement the median-type estimator for which most results were available and which was actually used in the simulation studies presented in Mondal and Percival (2012). In order to assess the estimators' performance, we test them in the same settings as those used in Section 4 of the main manuscript and we make use of the same measure of statistical performance which, as a reminder, is the Root Mean Squared Error (RMSE) defined as follows

$$\text{RMSE}^* := \sqrt{\text{med}\left(\frac{\hat{\nu}_j^2 - \nu_{j,0}^2}{\nu_{j,0}^2}\right)^2 + \text{mad}\left(\frac{\hat{\nu}_j^2}{\nu_{j,0}^2}\right)^2},$$

where $\nu_{j,0}^2$ represents the true model-implied WV for scale $\tau_j$. Figure F.1 represents the logarithm of this measure for all the considered processes and estimators. As can be observed, in the uncontaminated settings the best estimator is obviously the standard estimator CL which is however, in many cases, closely followed by the proposed estimator RWV while the alternative robust estimator MP is generally less precise/efficient than the other two estimators. In the contaminated settings however, the standard estimator becomes highly biased as expected while the two robust estimators are only marginally affected by the different forms of contamination. Between the latter two, the proposed estimator RWV is nevertheless the best since it reports a lower RMSE* for almost all considered scales of WV.

To conclude, the simulation study highlights how the proposed estimator RWV is the best alternative to the standard estimator CL in the uncontaminated settings while it is overall the best estimator in all the considered contaminated settings.

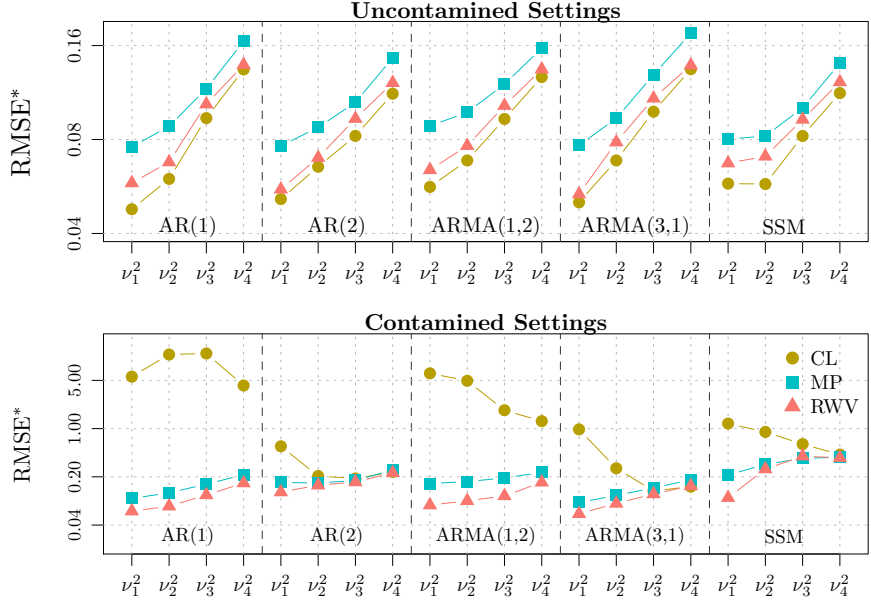

Figure F.1: Top row: RMSE* of the estimators of WV in an uncontaminated setting. Bottom row: RMSE* of the estimators of WV in a contaminated setting.

## F.2 Computational Efficiency

Aside from its statistical properties, another main point that this work aims to highlight is its computational efficiency. Indeed, the RGMWM can be computed in a fast manner which is particularly evident with more complex models, preserving this property as the sample size increases. Figure F.2 describes the median computational times (on a logarithmic scale) for different sample sizes for some of the models considered in the simulation studies in Section 4 in the main paper. In all cases, the computational times were taken within the uncontaminated sample setting and the sample sizes went from 100 to 10 million. In addition, the timing was stopped once each single estimation procedure reached, or went beyond, 6 hours. The computational advantage of the RGMWM (as of the GMWM) is evident from the plots in Figure F.2. Firstly, the ML is the most computationally efficient method for the **ARMA(1,2)** model estimation and is followed closely by the GMWM and RGMWM where their loss of speed compared to the ML is reasonable. However, if a fast and stable robust solution is needed, then the RGMWM is without doubt the best

28

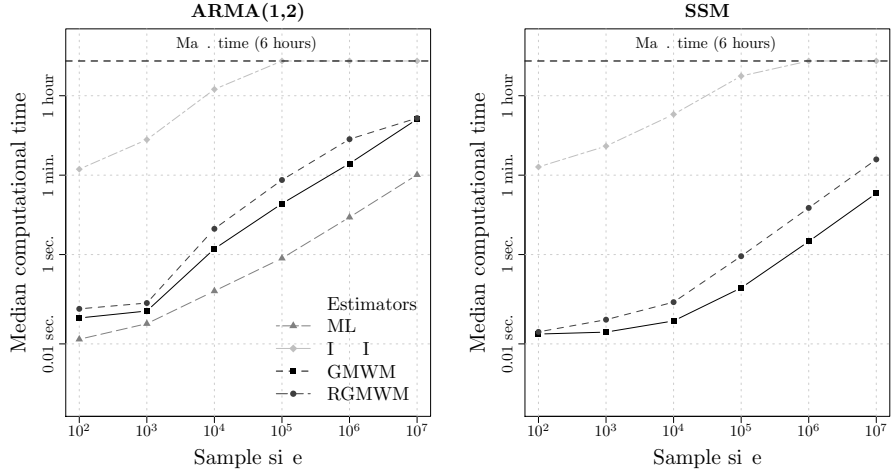

Figure F.2: Log-log plot of the median computational times (in seconds) of different estimation methods for different sample sizes for **ARMA(1,2)** (left) and **SSM** (right) models.

statistically-sound available approach. On the other hand, when considering more complex time series models such as the **SSM** then the RGMWM is by far the best robust solution from a computational speed standpoint. In addition, the RGMWM always converges in all the simulation settings considered while the indirect estimator (INDI) is unstable in various circumstances. Hence, not only can these models be computed quickly with the RGMWM, overcoming current limitations of alternative standard and robust methods, but they can also be estimated robustly without a considerable loss in computational efficiency.

# G   Additional Application: Inertial Sensor Stochastic Calibration

The data set in this section comes from the engineering domain and consists in the angular rate signal issued from a micro-electro-mechanical system gyroscope in static conditions. Due also to their low cost, these sensors are very common and are being increasingly used in the field of navigation engineering. The main goal of recording this kind of data is to improve the performance of the navigation sensors by identifying and estimating the parameters of the error model coming from the accelerometers and gyroscopes that

compose the sensor. Once these parameters are estimated they are inserted in a filter (usually an extended Kalman filter) which is used within a navigation system. The latter collects measurements from different sources such as Global Positioning Systems (GPS) or the inertial sensors themselves in an optimal manner in order to improve the navigation precision. Therefore, the latter greatly depends on the estimation of the parameters of the selected error model for the inertial sensor.

Figure G.3 shows the error signal from the gyroscope along with the outliers in a portion of the signal identified via the weights given to the observations by the RGMWM estimator. As can be observed, there are outliers that would appear to be obvious by simply looking at the plot and could be treated by fault detection algorithms for navigation systems (see further on) but there are many others that lie within the part of the signal which one would not expect to contain outliers. Despite the numerous outliers, these are extremely low in proportion to the whole dataset ($\approx 0.4\%$) which contains a little under 900,000 observations (issued from an approximately 2.5 hours-long recording sampled at 100 Hz). This may lead to think that estimations on this dataset would not be significantly influenced by outliers. Nevertheless, to understand how influential these observations could be, we estimated the classical and robust WV from the signal represented in Figure G.3. Using these estimates we then estimated an error model made by the sum of three latent first-order autoregressive models. This state-space model is among those suggested by Stebler et al. (2014) as being most appropriate to describe such signals. Table G.1 shows the estimated parameters for the GMWM and RGMWM estimators together with their confidence intervals (the ML was not considered for the same reasons given in Section 4 for the **SSM** model: the numerical stability and computational efficiency are unreasonable for this model and sample size). For both estimators the values of some autoregressive parameters are close to one, suggesting that the AR(1) model could be considered as a random walk. Indeed, a model that was commonly used to describe these signals was the sum of a white noise process with a random walk. However, Stebler et al. (2014) show how the use of sums of AR(1) models greatly improves the navigation performance over this model and the $J$-tests and confidence intervals support this view by ruling out the models which included a random walk. Although the differences between the estimations do not appear to be large since the estimated level of contamination is low, a significant difference is to be noticed for the parameters of the first two autoregressive processes indicating that the contamination appears to have an impact on estimation and that robust methods should be preferred (assuming the Gaussian assumption holds). Even one (or few) slightly misestimated parameter(s) can be highly relevant in the context of navigation systems since these are fed into the filters which will progressively misestimate the position as the sensors

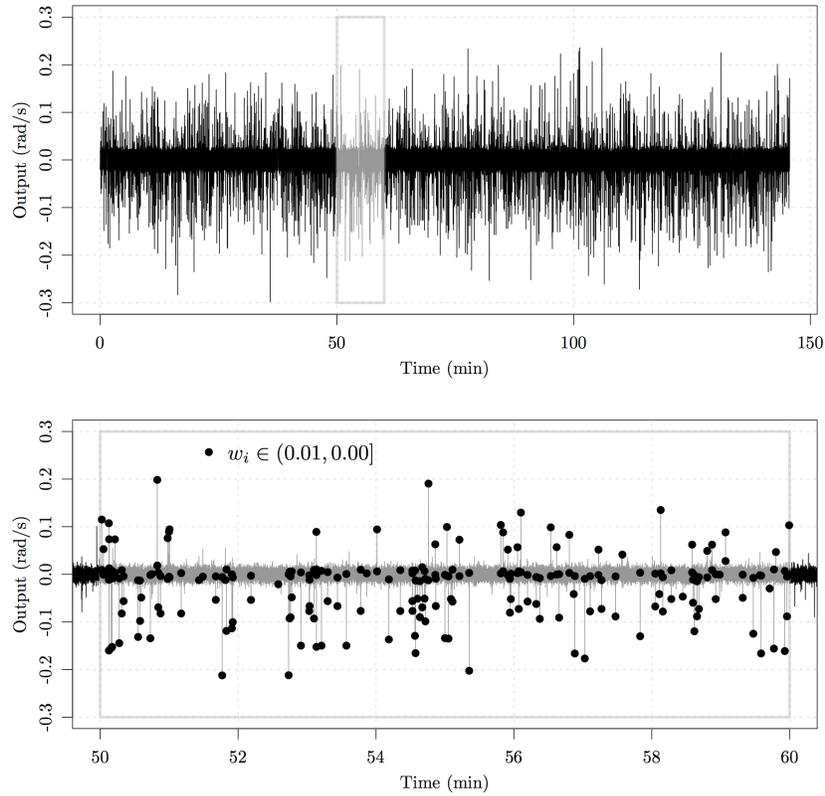

Figure G.3: Top part: Inertial sensor time series. Bottom part: zoom-in on grey part of the time series with black points indicating extreme outliers identified through the weights of RGMWM.

work in "coasting mode" (i.e. without the GPS integration) and deliver the so-called "error accumulation". Informally speaking this is due to the fact that these measurements are integrated several times and therefore their errors accumulate in time especially when no GPS observations are present to "reinitialize" the system (more details on this can be found, for example, in Titterton and Weston, 2004).

31

Table G.1: State-space model estimates for the gyroscope data in static conditions. Estimated parameters with GMWM and RGMWM estimators with $\rho_i$ being the $i^{th}$ autoregressive parameter, $\upsilon_i^2$ the innovation variance of the $i^{th}$ autoregressive model. Confidence intervals (CI) based on the approach used in Guerrier et al. (2013).

|  | GMWM | | RGMWM | |
|---|---|---|---|---|
|  | Estimate | CI($\cdot$, 95%) | Estimate | CI($\cdot$, 95%) |
| $\rho_1$ | $8.9546 \cdot 10^{-2}$ | ($8.9546 \cdot 10^{-2}$ ; $8.9546 \cdot 10^{-2}$) | $1.4816 \cdot 10^{-1}$ | ($1.4816 \cdot 10^{-1}$ ; $1.4816 \cdot 10^{-1}$) |
| $\upsilon_1^2$ | $7.8760 \cdot 10^{-5}$ | ($7.8612 \cdot 10^{-5}$ ; $7.8909 \cdot 10^{-5}$) | $5.5325 \cdot 10^{-5}$ | ($5.5211 \cdot 10^{-5}$ ; $5.5439 \cdot 10^{-5}$) |
| $\rho_2$ | $9.9831 \cdot 10^{-1}$ | ($9.9831 \cdot 10^{-1}$ ; $9.9831 \cdot 10^{-1}$) | $9.9687 \cdot 10^{-1}$ | ($9.9687 \cdot 10^{-1}$ ; $9.9687 \cdot 10^{-1}$) |
| $\upsilon_2^2$ | $2.5632 \cdot 10^{-10}$ | ($2.0887 \cdot 10^{-10}$ ; $3.0377 \cdot 10^{-10}$) | $1.0466 \cdot 10^{-9}$ | ($9.4291 \cdot 10^{-10}$ ; $1.1504 \cdot 10^{-9}$) |
| $\rho_3$ | $9.9997 \cdot 10^{-1}$ | ($9.9995 \cdot 10^{-1}$ ; $9.9999 \cdot 10^{-1}$) | $9.9997 \cdot 10^{-1}$ | ($9.9995 \cdot 10^{-1}$ ; $9.9999 \cdot 10^{-1}$) |
| $\upsilon_3^2$ | $1.1915 \cdot 10^{-11}$ | ($7.5299 \cdot 10^{-12}$ ; $1.6300 \cdot 10^{-11}$) | $1.3626 \cdot 10^{-11}$ | ($8.7412 \cdot 10^{-12}$ ; $1.8511 \cdot 10^{-11}$) |

Moreover, the proposed RGMWM can be of great usefulness in the area of Fault Detection and Isolation (FDI) for inertial measurement units (see for example Guerrier et al., 2012, and references therein) as shown in Figure G.3. In general, the task of FDI includes the detection of the presence of failures (or outliers) and the isolation of the component responsible of the irregularity. In the inertial navigation framework, FDI algorithms are used, for example, to ensure the safety of aircrafts or robots which deeply rely on inertial sensors. In fact, usual FDI methods in this area use various measurements coming from several sensors which entail a series of disadvantages. Moreover, these methods often make use of "rule-of-thumb" cut-off values which generally determine which observations are "unusual". On the other hand, the proposed approach would be able to detect "unusual" observations conditioned on the probabilistic behaviour of previous ones thereby determining more reasonable cut-off values. Although this is left for future research as some further adjustments would need to be put in place, our approach could be used as a basis for FDI by only using one signal coming from the sensor calibration procedure. One of the advantages of this approach, in addition to those already mentioned, is that it would have important impacts in terms of costs and constraints (e.g. weight, electric consumption, etc.) for robots or small unmanned aerial vehicles which are currently a major focus of technological and mechanical research.

}

\end{document}